\documentclass[journal,transmag]{IEEEtran}
\usepackage{tikz}
\usepackage{amsmath}
\usepackage{filecontents}
\usepackage[utf8]{inputenc}
\usepackage{verbatim}
\usepackage{amsmath}
\usepackage{graphicx}
\usepackage{subfigure}
\usepackage{multirow}
\usepackage{tabularx}
\usepackage{comment}
\usepackage{xspace}
\usepackage{makecell} 
\usepackage{algorithm}
\usepackage{algpseudocode}
\newcommand*\annotatedFigureText[4]{\node[draw=none, anchor=south west, text=#2, inner sep=0, text width=#3\linewidth,font=\scriptsize\sffamily] at (#1){#4};}
\newenvironment {annotatedFigure}[1]{
\centering
\begin{tikzpicture}
\node[anchor=south west,inner sep=0] (image) at (0,0) { #1};
\draw[red,thick,fill=red,fill opacity=0.2,draw opacity=0.2] (1.08,0.96) rectangle (2.6,3.6);
\draw[blue,thick,fill=blue,fill opacity=0.2,draw opacity=0.2] (5.32,0.96) rectangle (8.25,3.6);
\begin{scope}[x={(image.south east)},y={(image.north west)}]}{\end{scope}
\end{tikzpicture}}

\usepackage{amsmath,amssymb}
\usepackage{amsfonts}
\usepackage{bbm}
\usepackage{booktabs}
\usepackage{svrsymbols}
\usepackage{mathtools}

\DeclareMathOperator{\EX}{\mathbb{E}}

\newcommand{\RP}{\text{RP}}
\newcommand{\rp}{\text{ref}}

\newcommand{\noieee}{} 

\usepackage[normalem]{ulem}

\newcommand{\ZH}[1]{\textcolor{blue}{#1}}

\newcommand{\TPCvOne}{\text{TPCv1-offline}}

\newcommand{\threeradio}[0]{\emph{3-radio}}
\newcommand{\tworadio}[0]{\emph{2-radio}}

\DeclareMathOperator{\argmax}{\text{argmax}}

\DeclareMathOperator*{\AP}{\text{AP}}

\newcommand{\elevenk}[0]{\textit{11k}}

\algdef{SE}[DOWHILE]{Do}{doWhile}{\algorithmicdo}[1]{\algorithmicwhile\ #1}%
\algnewcommand{\algorithmicforeach}{\textbf{for each}}
\algnewcommand\algorithmicbreak{\textbf{break}}
\algnewcommand\Break{\algorithmicbreak}
\algdef{SE}[FOR]{ForEach}{EndForEach}[1]
  {\algorithmicforeach\ #1\ \algorithmicdo}
  {\algorithmicend\ \algorithmicforeach}
\algtext*{EndIf}
\algtext*{EndForEach}
\algtext*{EndWhile}

\begin{document}
\title{\Large \bf User-aware WLAN Transmit Power Control in the Wild}
\author{
\IEEEauthorblockN{Jonatan Krolikowski\IEEEauthorrefmark{$\ast$}, Zied Ben Houidi\IEEEauthorrefmark{$\ast$}, and Dario Rossi}
\IEEEauthorblockA{Huawei Technologies Co. Ltd, 92100 Boulogne-Billancourt, France}

\thanks{\IEEEauthorrefmark{$\ast$}{The first two authors contributed equally to this work.}}}

\IEEEtitleabstractindextext{%
\begin{abstract}
In  Wireless Local Area Networks (WLANs), Access point (AP) transmit power influences 
(i) received signal quality for users and thus user throughput, (ii) user association and thus load across APs and (iii) AP coverage ranges and thus interference in the network. Despite decades of 
academic research, transmit power levels are still, in practice, statically assigned to satisfy uniform coverage objectives. Yet each network comes with its unique distribution of users in space, calling for a power control that adapts to users' probabilities of presence, for example, placing the areas with higher interference probabilities where user density is the lowest. Although nice on paper, putting this simple idea in practice comes with a number of challenges, with gains that are difficult to estimate, if any at all. This paper is the first to address these challenges and evaluate  in a production network serving thousands of daily users the benefits of a user-aware transmit power control system.
Along the way, we 
contribute a novel approach to reason about user densities of presence from historical IEEE 802.11k data, as well as a new machine learning approach to impute missing signal-strength measurements. Results of a thorough experimental campaign show feasibility and quantify the gains: compared to state-of-the-art solutions, the new system can increase the median signal strength by 15dBm, 
while decreasing airtime interference at the same time. This comes at an affordable cost of a 5dBm decrease in uplink signal due to lack of terminal cooperation.
\end{abstract}
}

\maketitle


\section{Introduction}

Today more than ever, rich telemetry fuels data-driven network algorithms~\cite{jiang2017unleashing} --from CDN configuration~\cite{naseerconfiganator},  congestion control~\cite{dong2018pcc}, management of quality of experience~\cite{jiang2017pytheas}-- with the general idea to  sense the environment, learn its behavior or state in a given timescale, and take actions that drive the network toward a desired target.


In this context, IEEE 802.11 Wireless Local Area Networks (WLANs) are no exception. WLAN is almost the de facto access technology connecting users across home, campus and corporate environments:  a fleet of wireless Access Points (APs) are deployed to cover the physical space and connect mobile stations (STAs). 
Pioneering work suggesting to adopt a data-driven approach to control these notoriously chaotic~\cite{akella2005self} networks date back to over a decade ago~\cite{broustis2009measurement,5506716}. With more widely available and richer network telemetry, many aspects of WLANs are nowadays governed by  measurement-based control paradigms. A successful example is the auto-reallocation of channel and bandwidth resources to accommodate shifts in AP load profiles, as done for instance by Meraki's TurboCA~\cite{10.1145/3131365.3131398}. 


\begin{figure*}
    \includegraphics[width=2\columnwidth]{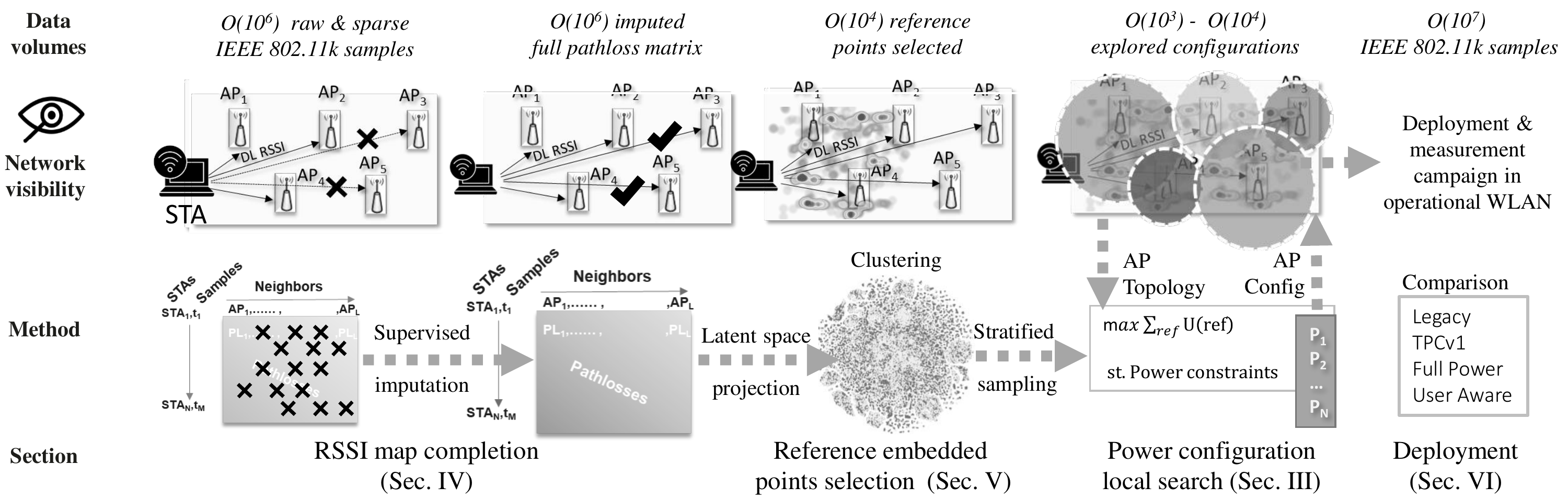} 
    \caption{Overview of the system and methods}
    \label{fig:architecture}
\end{figure*}

However, a crucial aspect of WLAN configuration, namely the transmit power levels of APs, has yet to fully explore data-driven opportunities.
Moreover, despite having been extensively studied in academia, mostly using theory or simulations (see Sec.~\ref{sec:related}), no sophisticated approach has made it to the industry, whose power control is still based on simple heuristics. 
Currently, network administrators often choose between two static and user-blind strategies, favoring one of two coarse and conflicting objectives of \emph{increased coverage} (good signal quality) vs \emph{reduced interference} (smaller contention probability).
One example is Cisco's Transmit Power Control, which offers two configuration policies (named TPCv1 and TPCv2). 
The first (v1) iteratively calculates transmit power levels for each AP, such that each AP is ultimately well covered, satisfying some signal strength quality constraint from its top-3 neighbors: as this obviously creates coverage overlap, it is likely to introduce more interference. To counter this, the alternative policy (v2) takes a conservative approach and, assuming that all APs share the same channel, sets out to minimize such overlap  creating coverage areas that are as disjoint as possible. The choice between the two policies is left to the 
the administrator -- although Cisco recommends to use v2 \emph{``only in cases where interference issues cannot be resolved by using v1''}~\cite{cisco_v1}.  

Yet, the above strategies only leverage inter-AP topology, while transmit power control could benefit from STA-level information such as user densities of presence. 
For example, if an area becomes physically inaccessible or rarely accessed (e.g., former office space converted into storage room), then  we might want to  decrease the transmit power to not cover it.
Similarly, it is reasonable to allocate more radio resources to areas with significant user presence, while still providing basic coverage but reducing interference to areas that historically show a lower user density.
Moreover, when such an area becomes accessible again or more frequently accessed (e.g., storage room converted back in office space), the configuration process should not incumb to the network administrator, but rather be data-driven (and either triggered by the admin, or even autonomously actuated by the network). 

However, user density of presence is difficult to directly gather from current telemetry. Intuitively, IEEE 802.11k (from here on: \elevenk) neighbor 
data allows to report for each connected user the  measured \emph{path loss} (PL) 
towards neighboring APs, which is an indication on users' points of presence. The latter has  proved promising to perform indoor localization~\cite{khalajmehrabadi2016joint, ayyalasomayajula2020deep}. Using \elevenk\ data to infer user density and optimize resources accordingly comes however with a number of challenges. First, unlike controlled settings, real-network measurements contain for each station PLs towards a handful of neighbors only. Second, assuming the full map is available, it is not clear how to assign resources proportionally to points of presence, let alone how to properly select the target points on which to optimize the network. 
Finally, assuming we solve all these challenges, even though nice on paper, it is unclear if user-awareness gives any benefit in realistic scenarios.

Addressing the above challenges, this paper is the first to add user-aware\-ness into WLAN power control and evaluate its benefits.
Summarizing our contributions: 

\noindent ($i$) We propose a new perspective to reason about user densities of presence, leveraging \elevenk\ neighbor data. Beyond power, our resulting user-aware utility 
(Sec.~\ref{sec:utility:definition}) can be easily extended to 
actions such as channel and bandwidth allocation.

\noindent ($ii$) Whereas state-of-the-art deep learning based imputation methods come short, our new multi-model approach, a necessary side-product of our system, successfully imputes missing signal-strength measurements with bounded error (Sec.~\ref{sec:impute:approach}). 

\noindent($iii$) We design and end-to-end system and perform a thorough experimental campaign showing feasibility and quantifying the gains: compared to state-of-the-art solutions, the new system can increase the median signal strength by 15dBm with no impact on interference. This comes at the cost of some limitations that we summarize in Sec.~\ref{sec:limit}.

In the remainder,  we overview our system (Sec.~\ref{sec:overview}) and separately evaluate its  building blocks  (Sec.~\ref{sec:powerls}--\ref{sec:refpoint}). We then report our results from a real deployment (Sec.~\ref{sec:results}). We finally place our work in the context of prior work (Sec.~\ref{sec:related}), summarize the limitations (Sec.~\ref{sec:limit}) and conclude (Sec.~\ref{sec:conclusions}).

\section{System Overview}
\label{sec:overview}
We overview the system at a glance with the help of Fig.~\ref{fig:architecture}.
The picture shows the input data and, from left to right, the required processing steps for real deployment. 
Annotations report the volume and type of data at each step, pointing to the relevant sections in the paper where the block is detailed.
\begin{figure*}[t!]
    \subfigure[Illustration of neighbor measurements]{%
        \includegraphics[width=.39\columnwidth,trim=0 -1cm 0 0, clip]{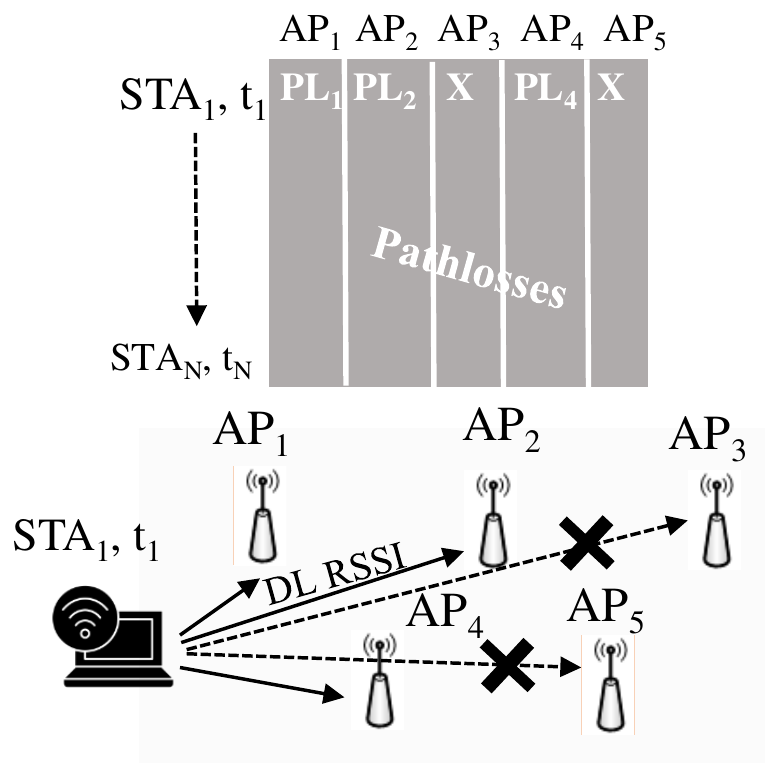}
        \label{fig:neigh:example}
    }
    \subfigure[Number of neighbors per measurement]{%
        \includegraphics[width=.79\columnwidth]{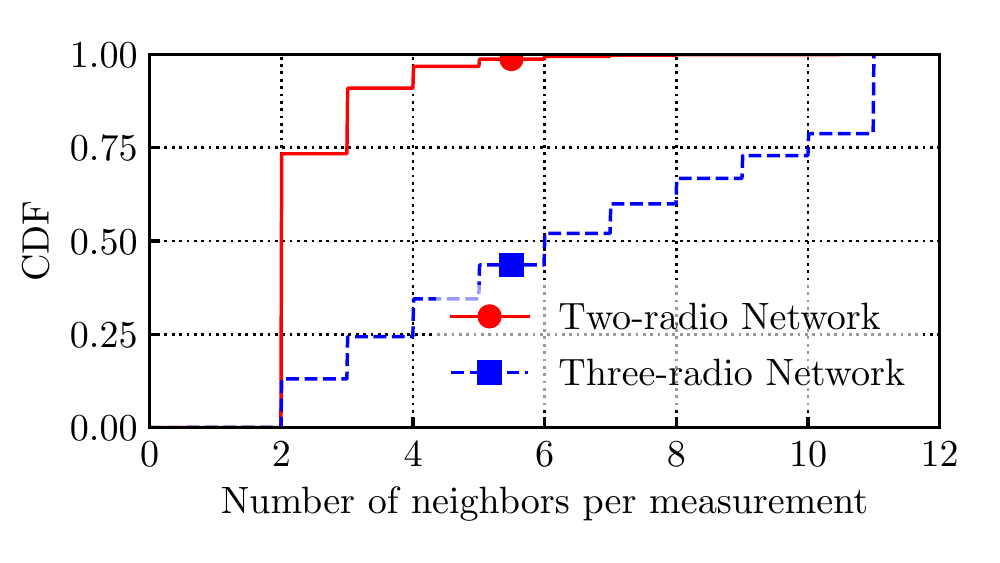}  
        \label{fig:num:neigh}
    }
    \subfigure[Path loss distributions]{%
        \includegraphics[width=.79\columnwidth]{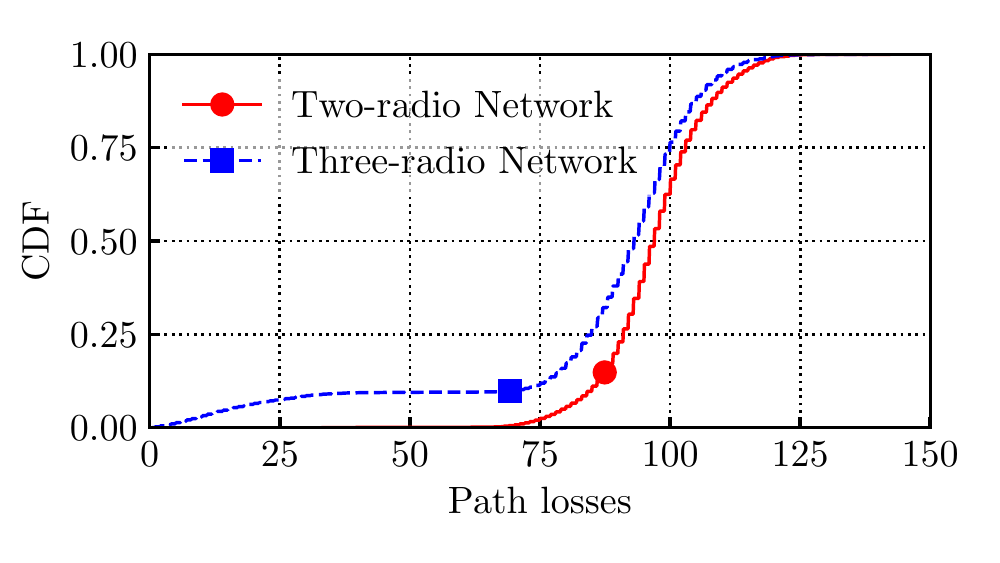}  
        \label{fig:pl:dataset}
    }
    \caption{IEEE 802.11k neighbor measurements overview: we obtain data from two distinct networks of 33 and 38 APs each. The first only has two service radios, the second has a third radio dedicated to measurements.}
\end{figure*}
More concretely, to infer densities of user presence in the network on a coarse timescale (e.g., weekly or monthly), we leverage historical \elevenk\ neighbor data from  operational deployments (see Sec.~\ref{sec:11k}). Once processed and aggregated, the data contains information about the PL between STAs and their neighboring APs, which we define as the measured signal attenuation or the difference between transmitted and received power.   
Rather than attempting to map measurements to exact physical coordinates, we argue that operating in a latent space preserving distance metrics is equally 
fit for the purpose of \emph{user densities estimation} which, to the best of our knowledge, we are the first to explore. 
User-awareness is attained via \elevenk\ measurements,
where users' past locations are abstracted as \emph{reference points} (RPs), i.e. a set of past \elevenk\ measurements of STAs, 
 having each a PL estimation towards all APs.
In essence, the assumption is that user habits are stable. Hence, popular latent positions, i.e., in the PL space, gathered from \emph{past} measurements will be also representative of popular latent positions in \emph{future}. 
In case of substantial environment or user behavior change, estimated densities must be updated with new measurements. In case of need, drivetests could be performed to augment user traces, making sure all relevant locations are covered.

We propose to optimize the network according to past \elevenk\ measurements, and gauge the quality of our optimization during a future deployment.  However, there are two practical reasons for which \emph{raw} data cannot be leveraged directly.
First, raw measurements are \emph{excessively sparse}: even in large and dense networks, the majority of samples contain RSSI measurements to very few neighbors. We find that sparsity can lead to severe bias in optimization, which mandates RSSI map completion (Sec.~\ref{sec:imputation}).
Second,  whereas the data is sparse (column-wise, according to representation of Fig.~\ref{fig:architecture}), the overall volume is \emph{excessively large} (i.e., row-wise), as it can reach millions of samples in a single day: thus, it is necessary to downsample the data to a computationally manageable amount (Sec.~\ref{sec:refpoint}). Whereas a \emph{uniform random sampling}  selects areas proportionally to their population (i.e., areas for which the 
PL vectors are similar for a large number of measurements), there is no guarantee that the selected RPs include less popular areas, 
hence negatively affecting coverage. Thus, we propose to cluster the latent PL space for \emph{stratified random selection},
to ensure it being \emph{coverage-friendly}. The selected RPs can be seen as an \emph{augmented trace} that we use to represent real user densities, and on which our trace-driven simulator will be applied.

Once RPs are selected, we employ a local search algorithm (Sec.~\ref{sec:powerls}) to explore power configurations that, given a channel allocation~\cite{iacoboaiea2021real},  maximize network quality, as described via a utility function.  
Finally, we apply the full RSSI map completion, 
RP selection and optimization pipeline in a real deployment, that we contrast to several alternatives in a thorough experimental campaign of several months (Sec.~\ref{sec:results}).
In the remainder of this section, we  detail 
the WLAN deployments and the \elevenk\  data (Sec.\ref{sec:11k}), as well as the definition (Sec.
\ref{sec:utility:definition}) and estimation (Sec.
\ref{sec:utility:evaluation}) of the utility function.
\subsection{IEEE 802.11k neighbor measurements}\label{sec:11k}
\paragraph*{IEEE 802.11k} Our system is data-driven, mainly fueled by \noieee
\elevenk\  neighborhood measurements. 
Originally, \elevenk\ data was meant to assist the handoff process: as the list of signal measurements towards the best APs is available to each STA, it allows for moving STAs to proactively roam to the best neighbor. 
More generally, such data allows to acquire a fine-grained user-centric view on network performance indicators, including neighborhood signal, channel load and STA-specific metrics such as downlink RSSI.  As illustrated in Fig.~\ref{fig:neigh:example}, each second, for each STA, we collect the signal strength (and subsequently the PL) towards neighboring APs.  While our \elevenk\ data do not carry physical positions of users (e.g. physical coordinates) they still express a distance (in the PL space) from STAs to potentially each AP in the network. 

\begin{table}[!t]
\caption{Summary of \elevenk\ \textit{neighbor data}.}\label{tab:datasets}
\begin{tabular}{ccccc}
\toprule
Network    & Duration (Period)                & Samples & AP & STAs/day \\ 
\midrule
\threeradio  & 6d  (Jul'21)   & 12.7M     & 38 &    4K     \\ 
\tworadio$^\dagger$  &   24d (May-Jun'21)    & 6.8M     & 33 &   2K    \\    
\bottomrule
\multicolumn{5}{l}{$^\dagger$ Results in Sec.\ref{sec:results} have been gathered in Sep'21-Jan'22.}
\end{tabular}
\label{sec:data:networks}
\end{table}

\paragraph*{Real network deployments}
Throughout the paper, we rely on two operational networks, summarized in Tab.~\ref{sec:data:networks}. 
We denote as \tworadio\ network a production WLAN deployment of 33 APs 
implementing 802.11ac, 
each having two radio interfaces (2.4GHz and 5GHz bands). 
 For this network, we 
are able to both \emph{measure}  (during May'21-Jun'21) 
as well as  \emph{actuate} (Sep'21-Jan'22).
In the \threeradio\ network, each of the 38  APs (802.11ax) is equipped with an additional 5GHz radio interface that is not used for payload transmissions, but is dedicated to measurements, specifically by  scanning  different channels. Here, we are only able to \emph{measure} (Jul'21), but not actuate. In the remainder, we focus on the 5GHz band, but our methodology applies regardless.

\paragraph*{Experimental measurements}
Statistical properties of \elevenk\ measurement collection are reported in  Fig.~\ref{fig:num:neigh}-\ref{fig:pl:dataset}.
In practice, the completeness of these measurement reports varies, and specifically depends on the availability of radio resources -- as APs in the 2-radio network need to switch to the STAs' channel before performing the measurement.  As shown in Fig.~\ref{fig:num:neigh}, 2-radio network scans 
are sparse and the number of APs per report is quite unimodal : 3/4 of measurements report the PL to just one additional AP on top of the AP to which the STA is connected, and 99\% report no more than 6 APs  across all measurements. While the 3-radio network measurements are more complete (median of 6 APs), RSSI map completion imputation remains necessary even here.

The corresponding PL distribution is reported in Fig.~\ref{fig:pl:dataset}: notice that whereas the number of APs per \elevenk\ report is higher in the 3-radio network (so that one would expect more faraway APs, hence a higher PLs, to be included), the report is instead biased toward closer APs which are more useful from a STA perspective for roaming purposes (so that measurements appear biased toward lower PLs). 

Overall, whereas 3-radio WLANs provide richer neighbor information, 2-radio networks are still prevalent as operators may be reluctant to dedicate a third radio entirely to measurements. Thus, we deploy our system in the 2-radio network as a conservative test. We expect the use of \elevenk\ data to be even more beneficial in the 3-radio case.

\subsection{A user-aware utility function}
\label{sec:utility:definition}
The aim of WLAN transmit power optimization is maximizing the quality of the WLAN as perceived by as many STAs as possible.
Operationally, as conceptualized by Ahmed and Keshav~\cite{ahmed2006successive} for physical points in the space, this translates into three goals: (i) \emph{maximizing signal strength}, since better signal allows for higher throughput during transmission, (ii) \emph{minimizing interference}, since interference increases waiting times and (iii) \emph{balance load among APs}, since overloaded APs need to split their resources among too many STAs. 

\paragraph*{\elevenk\ sample-level utility}
We accordingly define the utility to summarize these partly conflicting goals, e.g., increasing signal increases coverage but also increases interference.  For user awareness,  we use RPs as a substitute for 
physical points in the space,
and define a function $U_\rp$ accounting for the effects (i)--(iii) that  a given AP transmit power configuration $\mathbf{p} = (p_1,\ldots,p_{|\AP|})$ 
 has on an \elevenk\ RP $\rp$  as:
\begin{align}
    U_{\rp}(\mathbf{p}) = \frac{ \text{SIG}_{\rp}(\mathbf{p}) }{  \lambda_{AP_\rp}(\mathbf{p}) + \text{I}_{\rp}(\mathbf{p}) },
    \label{eq:utility:rp}
\end{align}
\noindent where $\text{SIG}_{\rp}(\mathbf{p})$ is a measure for the received signal strength at $\rp$, $AP_{\rp}(\mathbf{p})$  is the AP that $\rp$ associates to under $\mathbf{p}$,  $\lambda_{{AP}_{\rp}}(\mathbf{p})$ its load,   and $\text{I}_{\rp}(\mathbf{p})$ the interference experienced by $\rp$ under configuration $\mathbf{p}$. Note that $\text{I}_{\rp}(\mathbf{p})$ depends on the channel configuration implemented in the network, which we assume as given, and account for it~\cite{iacoboaiea2021real}.

\paragraph*{Network-wide utility}
The definition (\ref{eq:utility:rp}) is in line with our goals, since it is (i) proportional to signal strength and inversely proportional to (the sum of) (ii) interference and (iii) load.  We then next define the overall utility of configuration $\mathbf{p}$
for the whole network
as the sum of utilities over the set of selected points $\RP$, as:
\begin{align}
\label{eq:util}
    U(\mathbf{p}) = \sum_{\rp\in\RP} \log(U_{\rp}(\mathbf{p})).
\end{align}

\noindent The logarithm is expected to provide fairness over RPs by overemphasizing those with weaker utility and avoiding few RPs with large utility to have a dominant impact. 

\subsection{Estimating the utility}\label{sec:utility:evaluation}

\paragraph*{Options} Naïvely, given a power configuration, two options are available to compute signal strength, load and interference at a given RP (cf.  Eq.\eqref{eq:utility:rp}). Roughly, one can (i) apply the configuration directly on the \emph{real target network}, and measure the impact on the RP, or (ii) 
estimate them via an \emph{approximated digital replica}  of the network: a simulator or numerical solver. 
The first is virtually impossible if RPs come from past real user traffic. Besides, we would need to measure utility for thousands of RPs, and all possible configurations.
We instead search for the best solution after exploring, offline in a simulator, a large fraction of the configuration space.
Clearly, there is a trade-off between the simulation feasibility, realism and speed of computation. But since our exploration is not based on gradients, the required level of realism is low, i.e. just being able to compare two solutions.
Moreover, as the simulator is used only in the search 
whereas performance is evaluated in real conditions,  the simulator's lack of fidelity is not reflected in the reported results.

\paragraph*{Simulator}  We opt for a simplified trace-driven simulator in which the received signal strength at an RP or STA\footnote{We remind that an RP corresponds to a STA in a given point in time.} is computed as the difference of the serving AP transmit power and the respective PL. The serving AP is estimated based on the simplifying assumption that STAs connect to the  AP with the strongest signal. Then, AP load can be estimated assuming uniform STA load and, given channel configuration, AP interference can be estimated as well. 


As a preliminary, we evaluate the Pearson correlation between the simulated airtime interference and the measured one, based on over 20 weeks of distinct power configurations in our \tworadio\ 33 AP WLAN. 
We find a median correlation of $0.4$, and a p-value $<0.05$ for over 95\% of samples.
Given how noisy wireless environments are, we consider this to be satisfying.\footnote{An equivalent of 20 runs of random interference values in the same range obviously gives 0 as correlation.} Indeed, we do not want our utility to be the most accurate average interference predictor, but only to push the search algorithm, on average, in the right direction. This important result allows us to confidently state that the enhancements, which we see later in all our deployments, are due to our approach and not to some side-effect.


\begin{figure*}[t]
\centering
    \includegraphics[width=.99\textwidth]{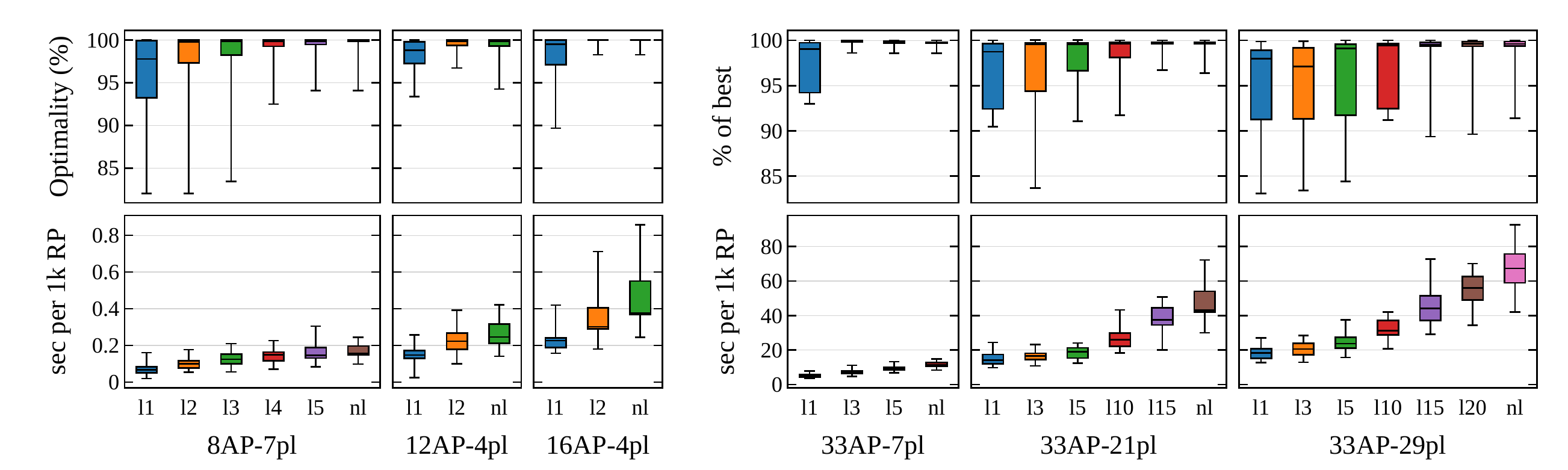}  
    (a)$\qquad\qquad\qquad\qquad\qquad\qquad\qquad\qquad\qquad\qquad$(b)
    \caption{Performance of our LS (Algorithm~1) across several topologies and power-level ranges. (a) Optimality gap for toy case scenarios of 8-16 APs and 4-7 power levels
    where optimal solution is found via exhaustive search. (b) Gap with respect to the best found solution for realistic scale topology (33 APs) and 7-29 power levels.
    }
  \label{fig:powerLS}
\end{figure*}

\section{Power Optimization Algorithm}
\label{sec:powerls}
\subsection{Method}\label{sec:sim:necessity}

The utility function given in Eq.\eqref{eq:util} is  non-linear, and has numerous points of discontinuity. As such, finding the optimal power configuration with Mixed-Integer Linear Programming or Gradient-based methods  seems ill-advised. We hence resort to a local-search (LS) based heuristic -- that exhibits a low optimality gap in small scale scenarios (where the optimum solution can be found by exhaustive search), and a limited computational complexity for larger scenarios.

\paragraph*{Input} More specifically, the goal of LS is to find the best power configuration for each AP, given as input: (i) the discrete set of available transmit power levels for each AP, (ii) the channel configuration of each AP, (iii) a matrix of measured  PLs among AP pairs, and finally (iv) a matrix of measured PLs from a set of   RPs toward each AP. 



\begin{algorithm}[t]
\caption{Local search (LS)}
\label{alg:powerLS}
\begin{algorithmic}[1]
	\small
\State Input $\mathbf{p} = (p_{1}, \ldots, p_{|\AP|})$ \algorithmiccomment{arbitrary power level vector} 
\State $\mathbf{p}^{\ast} \coloneqq \mathbf{p}$\algorithmiccomment{initialize best solution}
\While{$\text{not interrupt}$}\algorithmiccomment{local optimum or time-limit}
    \ForEach{$a\in\AP$}\algorithmiccomment{iterate single nodes}
        \State \label{alg:step:optAP}{$p_{a}^{\ast}\coloneqq \argmax_{p} U(p, \mathbf{p}^{\ast}_{-a})$ \algorithmiccomment{best for $a$ among $L$ trials}}
    \EndForEach
    \State $\mathbf{\hat{p}} \coloneqq \argmax_{a} U(p_{a}^{\ast}, \mathbf{p}^{\ast}_{-a})$ \algorithmiccomment{best solution so far}
    \State $\mathbf{p}^{\ast} \coloneqq \argmax \lbrace U(\mathbf{\hat{p}}),  U(p_{1}^{\ast}, \ldots, p_{|\AP|}^{\ast})\rbrace$ \algorithmiccomment{speedup}
\EndWhile 
 \State \Return $\mathbf{p}^{\ast}$
\end{algorithmic}
\end{algorithm}


\paragraph*{Algorithm}  LS is detailed in Algorithm~\ref{alg:powerLS}. Starting form a randomly selected input power configuration, the algorithm iterates repeatedly over the APs, varying their power levels individually (line 5).
Note that the number of configuration trials per AP can be capped to an explicit limit $L$ to reduce the computational complexity of the search: each evaluation of the utility function Eq.\eqref{eq:util} implies a call to a relatively complex computation (Sec.~\ref{sec:utility:evaluation}) that needs to iterate  
over possibly several thousands of selected  (Sec.~\ref{sec:utility:definition}) RPs.
At each network-level iteration, the best found solution (line 6) is checked against the solution that combines the best individual power levels (line 7) for possible speedup. The search  terminates when no further improvement can be found\footnote{Note that the algorithm is guaranteed to terminate since the utility increases in each iteration, and there is a limited  set of available configurations.} or when a time cap is reached, in order to bound running time. 

\subsection{Evaluation}
\paragraph*{Toy case scenario}
We first resort to toy cases to precisely evaluate the optimality gap of LS
for cases where the optimal can be found via exhaustive search. 
Due to combinatorial explosion of the search space, we limit the topology size to 8-16 APs, and restrain power settings to 4-7 power levels. For each scenario, we evaluate performance over 32 random instances, starting from a randomly generated power configuration for each AP.
In each scenario, we explicitly control the extent of the algorithmic exploration for each AP by capping the limit of the number of per-AP trials $L$ in each step (line 5), and contrast it to an unbound number of trials (i.e., systematic exploration of all available power levels at each iteration, denoted as \emph{``nl''} for no limit).  

From the top plot of Fig.~\ref{fig:powerLS}-(a),  we observe that LS-nl reaches optimality (median across random instances) in all three scenarios. We further observe that in such toy settings, the optimality gap remains very small even with limited per-AP trials: notice that in 75\% of the cases, the optimality gap is less than 3\% for LS-l2 (for 4 power levels scenario) and LS-l4 (for 7 levels). 
The bottom plot of  Fig.~\ref{fig:powerLS}-(a) complements the picture by showing run time, expressed in seconds per thousand  RPs to be considered in the evaluation of the utility function (runtime is gathered on a single core of an Intel Xeon Platinum 8164 CPU at 2.00GHz). As expected, larger search translates into longer computation, with evident diminishing returns in terms of solution quality (e.g., for LS-nl). These results confirm the soundness of our algorithm.

\begin{figure*}[t!]
    \subfigure[Effect of missing STA-AP pathloss]{%
    \includegraphics[width=1.\columnwidth]{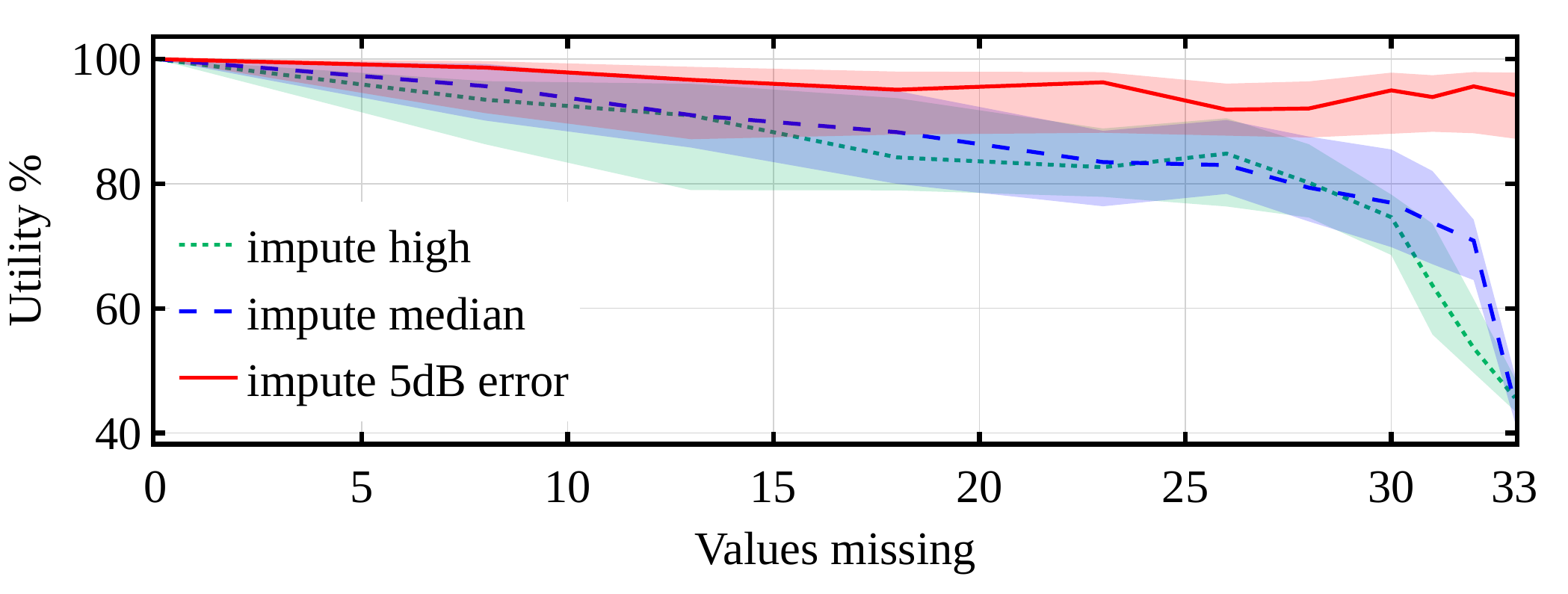}  
    \label{fig:impute:obfuscation}
    }
    \subfigure[ ML vs. baseline (\threeradio\ network)]{%
        \includegraphics[width=.99\columnwidth]{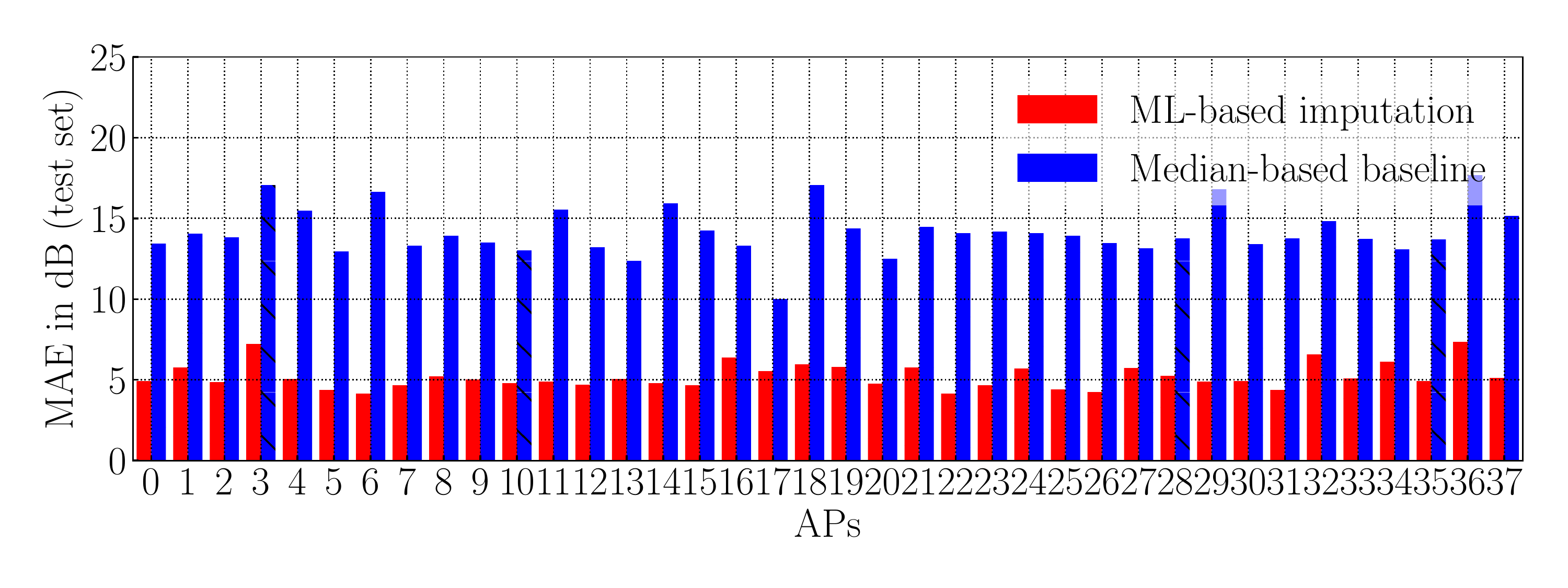}  
        \label{fig:impute:3r:varyimpute}
    }
    \caption{Necessity of imputation (a) and   machine learning based imputation (b)}
\end{figure*}

\paragraph*{Realistic scenario}
Next, we apply LS to larger instances, fit on the 33-AP \tworadio\ network, introduced earlier.  We use 32 random initial configurations, and restrict power settings to
9--15 dBm (7 power levels),  4--24 dBm (21 levels) and the full scale 4--32 dBm (29 levels).  The top plot of Fig.~\ref{fig:powerLS}-(b) shows in this case a comparison to the best solution found by running LS from many starting configurations for up to a day. Even in this more challenging scenario, we observe a small gap and diminishing returns: we hence fix our choice to LS-l15, that is well fit for finding a good enough solution in a real-scale scenario in less than 60sec per 1000 RPs.

\section{RSSI Map Completion}
\label{sec:imputation}

We first show why it is necessary to complete missing RSSI measurements, then design and evaluate a novel approach to impute missing data.

\subsection{Necessity}\label{sec:imp:necessity}

\paragraph*{Preliminaries} As shown in Fig.~\ref{fig:num:neigh}, \elevenk\ data provides a limited view of the AP neighborhood for each STA. We  first assess the impact that a limited horizon has on the network utility evaluation~Eq.\eqref{eq:util}.
For this purpose, we synthetically create a set of 32 random topologies of 33 APs, and select 330 RPs for each topology. For the sake of simplicity, we do so by placing APs and STAs randomly on a square, deriving  PL values from Euclidean distances with a simple PL model. 
We obtain a synthetic ground-truth corresponding to an ideal case with \elevenk\ PL  measurements from each STA to all APs (i.e. an unlimited horizon). For this ideal case,  we obtain the best configuration using the just-described LS  algorithm, and gather the corresponding utility that we use as a reference. 
We then limit the horizon of \noieee \elevenk\ measurement by systematically obfuscating some of the STA-AP PL pairs, specifically biasing obfuscation towards the furthest APs (as noted earlier in Fig.~\ref{fig:pl:dataset}).   We next complete the matrix in three increasingly sophisticated ways, to provide the imputed matrix as an input to LS, namely: (i) \emph{impute high},  assumes missing APs to be  beyond interference range and sets a very high pathloss; (ii) \emph{impute median}, is a simple data-driven strategy that learns the overall PL distribution and imputes missing values with the  median observed PL; (iii) \emph{impute 5dB error}, simulates a data-driven ML strategy that learns to impute missing-value per-AP but is slightly imperfect\footnote{In practice, we leverage the true PL that we distort with a mean absolute error of 5dB.} like ours.


\paragraph*{Impact}
We compare the utility degradation from 
full-visibility as a reference, to 
shortsighted \noieee \elevenk\ measurements imputed with strategies (i)-(iii) in Fig.~\ref{fig:impute:obfuscation}, 
for a variable number of missing values (curve reports the median, shade the 1st and 3rd quartiles).
When many true PL values are visible, the  performance degrades only little with any imputation scheme. When few true PL values are visible, losses can be significant for (i) high or (ii) median imputation: when only 2 (6) out of 33 PLs are available as typical for 2-radio (3-radio)  networks, the utility loss is 25\% (20\%).  Conversely, utility remains mostly unperturbed (<5\% loss) for a hypothetical ML regression able to estimate PL within 5dB from the true value.
Fig.~\ref{fig:impute:3r:varyimpute}  depicts  the PL estimation  (per-AP  mean absolute error) for the \emph{median} imputation strategy (blue bars, close to 15dBm error), and of \emph{our actual ML} strategy (red bars, close to 5dBm). This confirms the need for devising an accurate augmentation technique, that  we describe in the next section.



\begin{figure}[t]
    \includegraphics[width=\columnwidth]{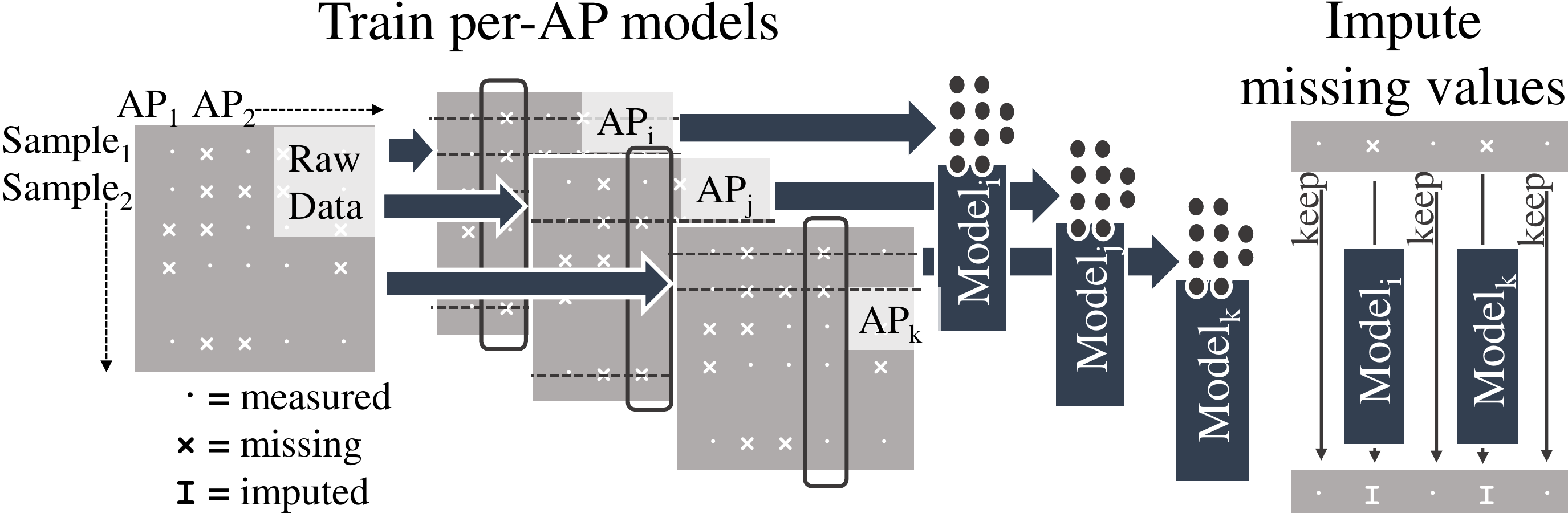}
    \caption{Approach for RSSI map completion.}
    \label{fig:imputation:ov}
\end{figure}

\subsection{Approach}\label{sec:impute:approach}
\paragraph*{Alternatives}
After initially trying them, observing poor performance,
we realized that state-of-the-art imputation methods are not applicable to our task: both the recent GAN-based imputation  MisGan~\cite{li2019misgan}, designed for image completion, as well as the well known Singular Value Thresholding (SVT)~\cite{cai2010singular} for matrix completion, rely on the assumption that values are missing  \emph{completely at random}. However, as shown earlier, PLs reported via \noieee \elevenk\ are biased towards closer APs (since useful for roaming) while high PLs tend to be missing. 

\paragraph*{Idea.} We thus design a practical ML-based approach for this problem. We  refrain from leveraging field tests on the target network, which would simplify ML models training and evaluation, but is of limited interest as inherently not portable, and hard to replicate for other networks.  To overcome the lack of groundtruth from field tests,   we design the novel approach presented in Fig.~\ref{fig:imputation:ov}. 
Note that, since the utility function computation only requires to learn \emph{distances}, we do
not need to perform a fine-grained \emph{localization}.
Further consider that a PL vector that contains only, e.g., 3 PL values towards 3 APs, is enough to roughly position the point in the latent PL space. Once this latent position is known, it is possible to learn a model for each AP to forecast the PL to each other point (i.e., each other AP or STA) in the latent space. As we do not need a precise  spatial position, we  simply train \#$\AP$ models, each  estimating the distance (in PL) towards one particular AP.

\paragraph*{Algorithm}  As illustrated in Fig.~\ref{fig:imputation:ov}, building a separate dataset to train (and validate) ML models for each AP is straightforward compared to building a full groundtruth: in practice, it is sufficient to take existing measurements in which the PL to the current AP is present, together with at least three other PL values toward other APs. 
For any given model $i$ trained for the $i$-th AP, the PL to the other APs is the input $x$,  and the  PL to the $i$-th  AP is the value $y$ to be learned. Said differently, we simultaneously create several datasets looping over each vector of \noieee \elevenk\ STA  measurements $x=(PL_{1}, …, PL_{|\AP|})$ where at least 4 PL values are not missing. We then  purposely ``hides'' the  measurement value $PL_{i}$ in the vector $x$ and uses this $PL_{i}$ as the true label for the $i$-th AP model.
By doing so over all APs and measurements, we automatically obtain a \textit{labeled dataset per AP}. 
Any supervised regression technique can then be used to learn the missing value. In our experiments, after further splitting each per-AP labeled data into training, validation and test sets, we converged to a simple 3-layered dense neural network with about 37k trainable parameters. 
As the procedure is fully automated and no manual labeling is needed, the technique can be seamlessly ported to different networks where \noieee \elevenk\ STA measurements are available.


\begin{figure}
    \subfigure[Absolute error (2-radio)]{%
        \includegraphics[width=.48\columnwidth]{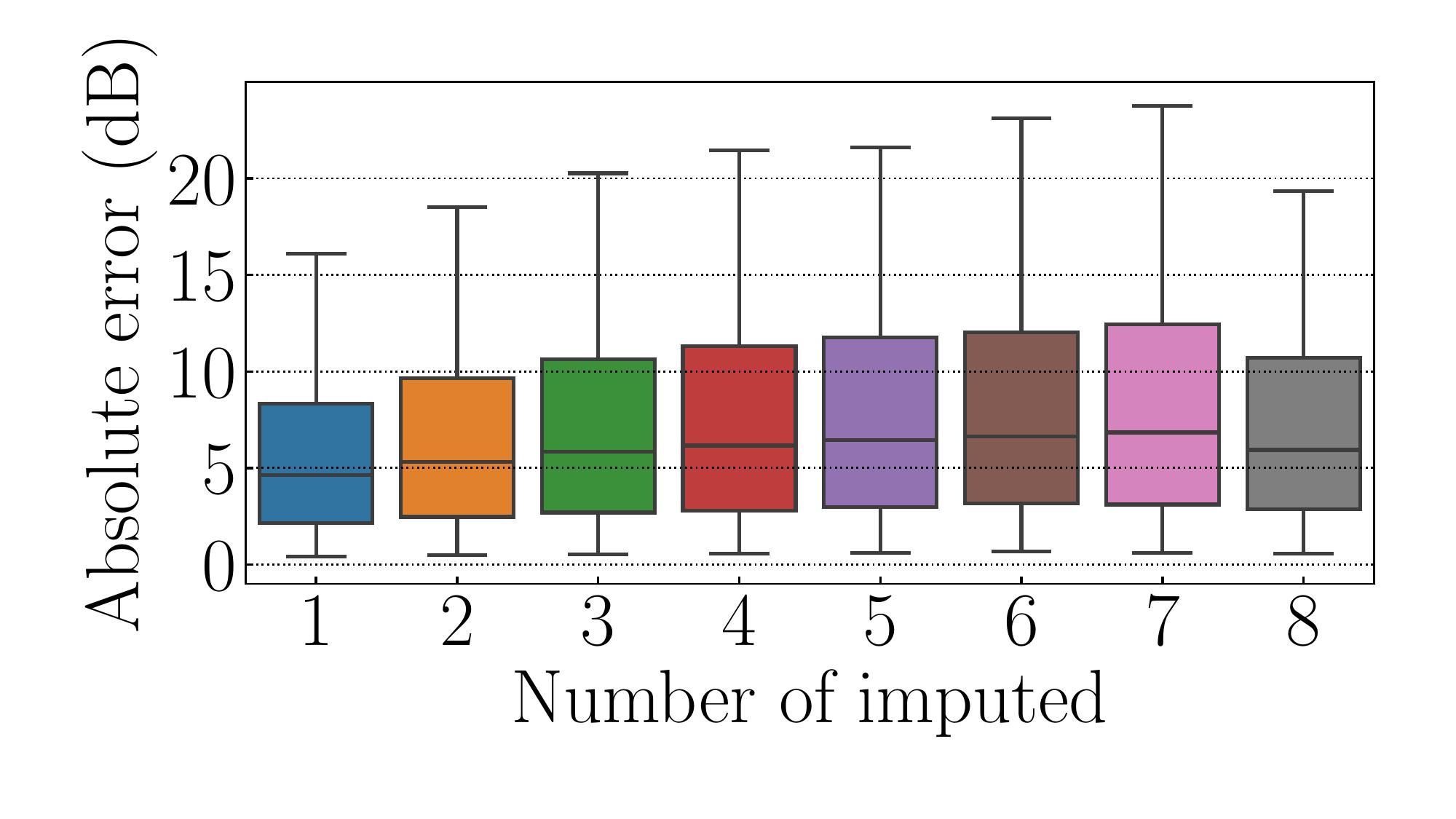}  
        \label{fig:2r:varyimpute}
    }
    \subfigure[ Absolute error (3-radio)]{%
        \includegraphics[width=.48\columnwidth]{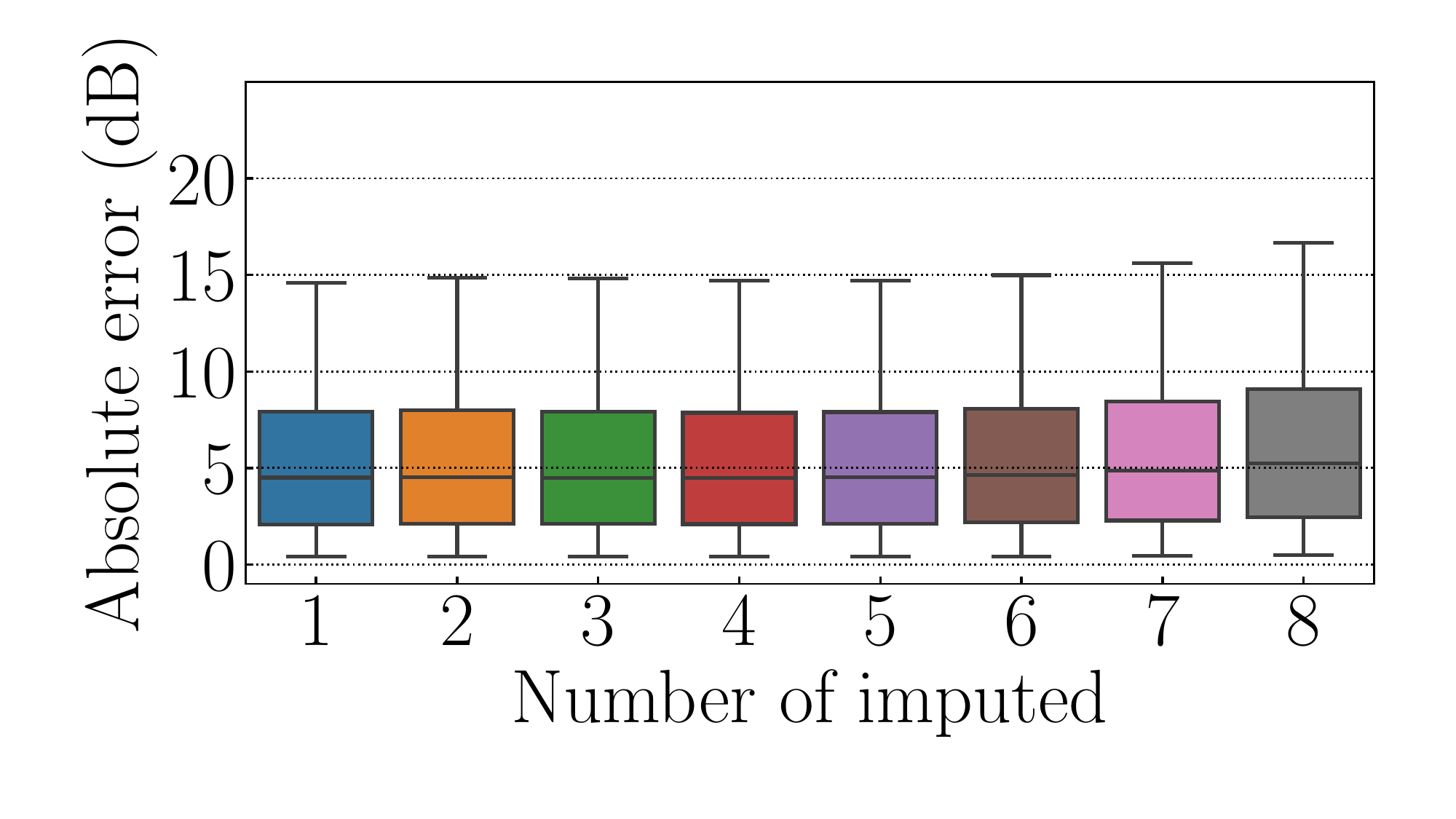}  
        \label{fig:3r:varyimpute}
    }
    \subfigure[Imputed vs. True (2-radio)]{%
        \includegraphics[width=.48\columnwidth]{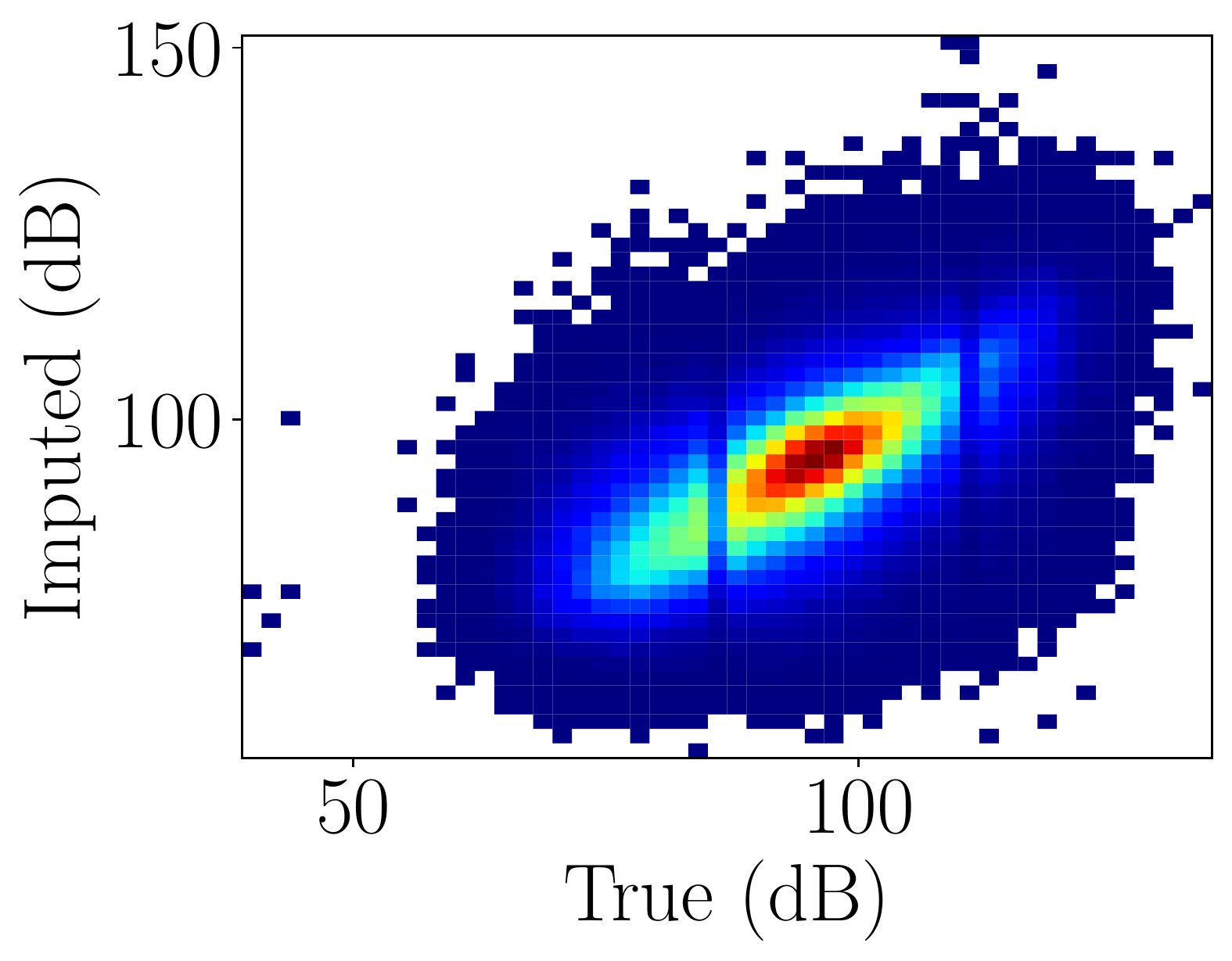}  
        \label{fig:scattertrueimp:2r}
    }
    \subfigure[Imputed vs. True (3-radio)]{%
        \includegraphics[width=.48\columnwidth]{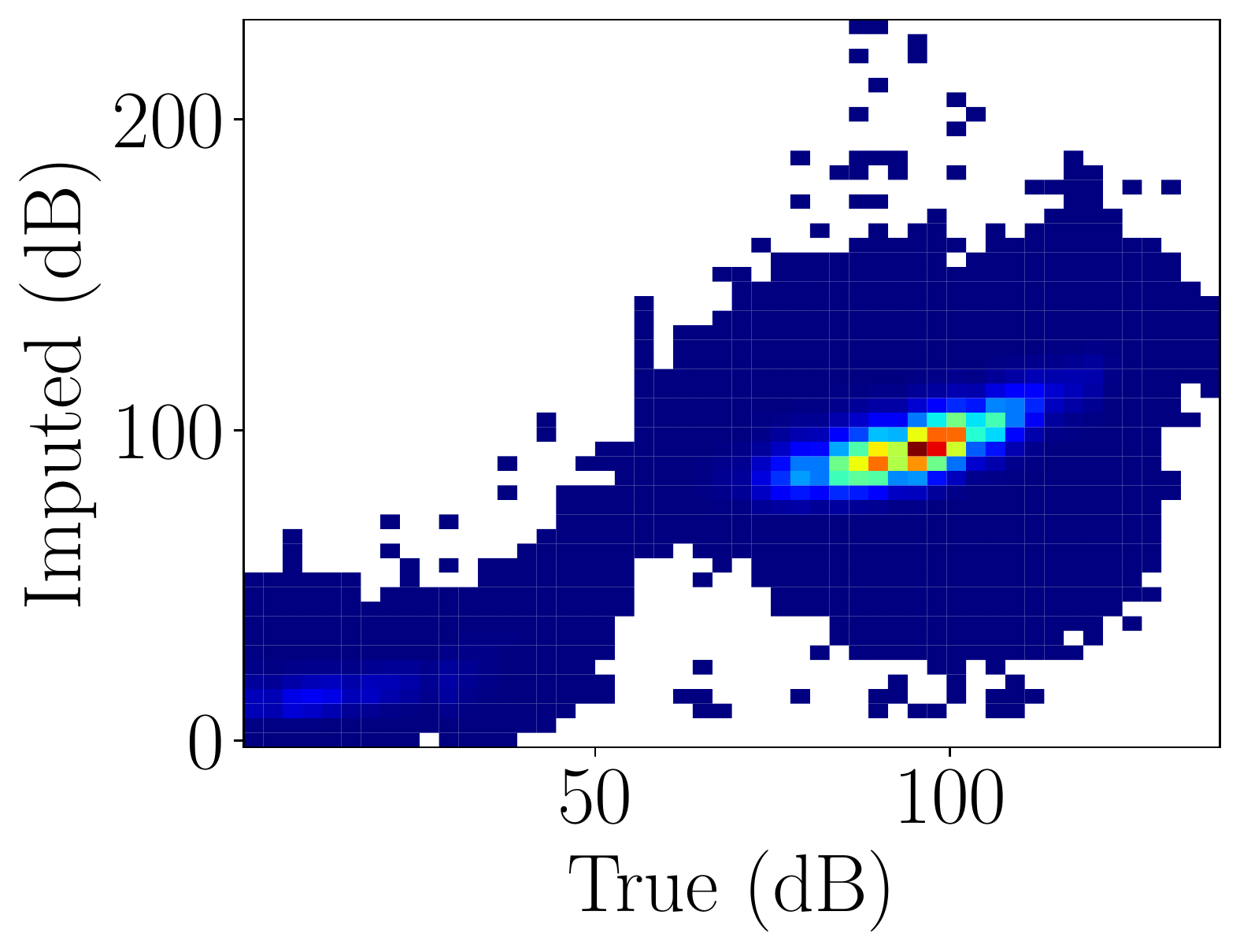} 
        \label{fig:scattertrueimp:3r}
    }    
    \caption{Imputation performance on hide-and-impute task for 2-radio and 3-radio networks.}
  \label{fig:impute:hide}
\end{figure}

\subsection{Evaluation}
\paragraph*{Preliminaries}  We now provide a more detailed evaluation of the machine learning (ML) imputation mechanism for the 2-radio and 3-radio environments introduced earlier in Tab.~\ref{tab:datasets}. Clearly, requiring multiple AP PL readings in the same \noieee \elevenk\ sample essentially filters out a fraction of the samples that are not useful for learning (i.e., those with less than 4 readings). In the case of the 2-radio network, only  a small fraction of the samples is thus exploitable, which still leads to 50k samples for training/validation per AP and 17k samples for test. In the case of the 3-radio, we use 1.5M samples for training/validation and 0.5M samples for test.  The number of samples per AP is roughly balanced,  with the exceptions of a couple of APs that gathered about 10 times fewer measurements than the rest.  As for the ML model, after preliminary investigation we resort to a rather shallow neural architecture with comprising 3 densely connected hidden layers of size 200/100/40, for overall 37k trainable weights. 

 
\paragraph*{Per-AP performance} 
We first evaluate our models separately on the original task on which they were trained, i.e. per-AP imputation, that we have briefly introduced in Fig.~\ref{fig:impute:3r:varyimpute}. The figure separately shows the performance on the test set for all per-AP models: compared to a statistical baseline (e.g., median imputation), the ML models drastically reduce the PL forecast error, ultimately limiting the utility function degradation in presence of missing values.

\paragraph*{Imputing multiple values} 
As \elevenk\ readings are sparse, to be useful for our purposes, ML-imputation needs    to successfully impute several missing values at once, as illustrated in the inference part of Fig.~\ref{fig:imputation:ov}. To stress test the model ability to do so in a controlled manner, we design a ``hide and impute'' task in which we hide more than one single AP at a time and use multiple ML models for imputation. 

We study in Fig.~\ref{fig:impute:hide} the effect of a variable number of imputed values: as  most of our \elevenk\ samples have less than 10 neighbors, this limits the boundaries of the controlled validation -- while far from the 33 and 38 APs that form the total number of APs in our datasets, the task already gives an   an idea of whether our approach would suffer depending on the number of values to impute.
Results show overall that, especially for the \threeradio, the performance is not affected by the number of imputed values, with a median error of 5dBm.
As expected, the \tworadio\ network is more challenging than the \threeradio\ one: nevertheless, even in this case only a moderate   degradation is observed, with median error increasing from 5dBm to about 7.5dBm when 8 values need to be imputed  (recall in the 2-radio case, the number of samples with more than 6 population is significantly smaller). As a complement, Fig.~\ref{fig:scattertrueimp:2r}-(d) shows the heatmap of true vs imputed values in both networks.


\section{Reference Point Selection}
\label{sec:refpoint}

\subsection{Necessity}\label{sec:ref:necessity}

Our objective is to use historical \elevenk\ data as an estimate of future user densities.  In order to cover all relevant STA positions, 
historical data needs to be down-sampled to be computationally manageable, while remaining representative.

\paragraph*{Complexity argument}
A na\"ive option is 
to leverage \emph{all} the data collected at such timescale for the purpose of optimization.  Recall however that in the WLANs of 
Tab.~\ref{tab:datasets}, the \elevenk\ data collection generates on average 283k (2.1M) samples/day in the 2-radio (3-radio) deployment. Since we need at least 3 measured AP PLs for imputation (see Sec.~\ref{sec:impute:approach}), we have, e.g.\ 1.8M usable samples in the 2-radio network. Computing the utility function leveraging all samples has then a prohibitive cost: recall that when aiming to a realistic scenario with 29 power levels, LS-l15 requires nearly a minute for 1k samples. Thus, the need to reduce computational complexity is a first reason for downsampling of \elevenk\ data.

\paragraph*{Representativeness argument}
A second, equally important reason is that the information content brought by each new sample is expected to have, at best, diminishing returns 
(e.g.\ since many samples come exactly from the same position) 
and in the worst case to induce a bias in the evaluation (e.g.\ as zones where STAs generate more samples likely get more importance in the optimization process).
Thus, a second major reason to  downsample the full RP set is to ensure representativeness of the selected RPs, e.g.\ by 
favoring performance in crowded areas vs  coverage of the whole  network area.

\begin{figure}[t!]
     {\hspace*{18pt}\centering\subfigure[Impact of radius on number of selected reference points]{%
        \centering
        \includegraphics[width=0.8\columnwidth]{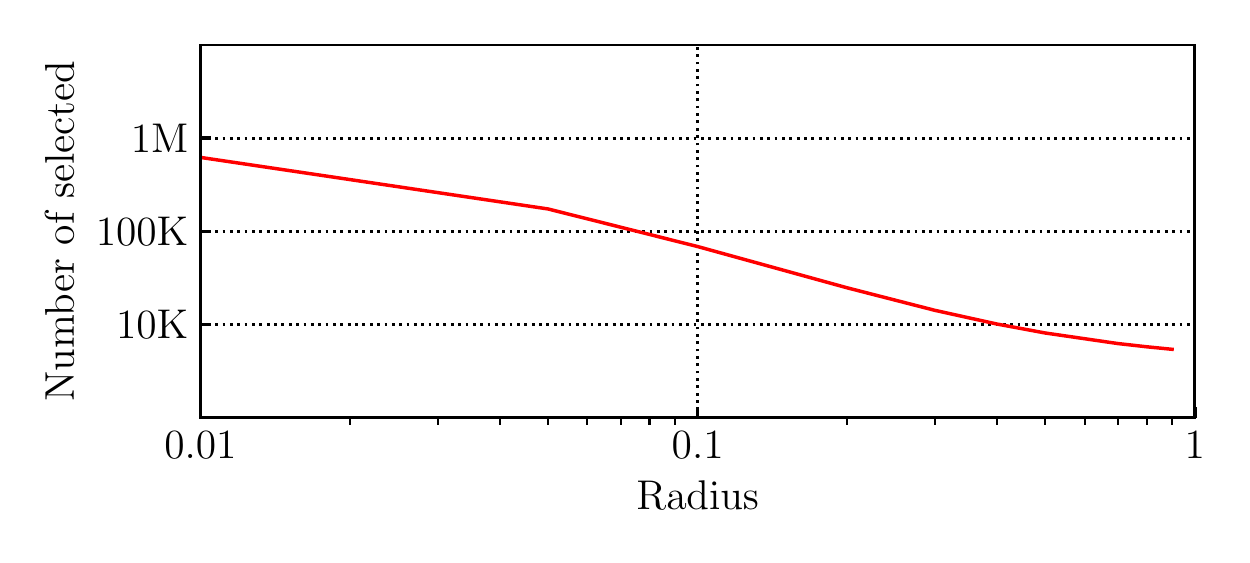}  
        \label{fig:radius:impact}
    }
    }
        {\centering
        \subfigure[Radius$=$0.1]{%
            \includegraphics[width=.45\columnwidth,trim=4cm 4cm 4cm 4cm,clip]{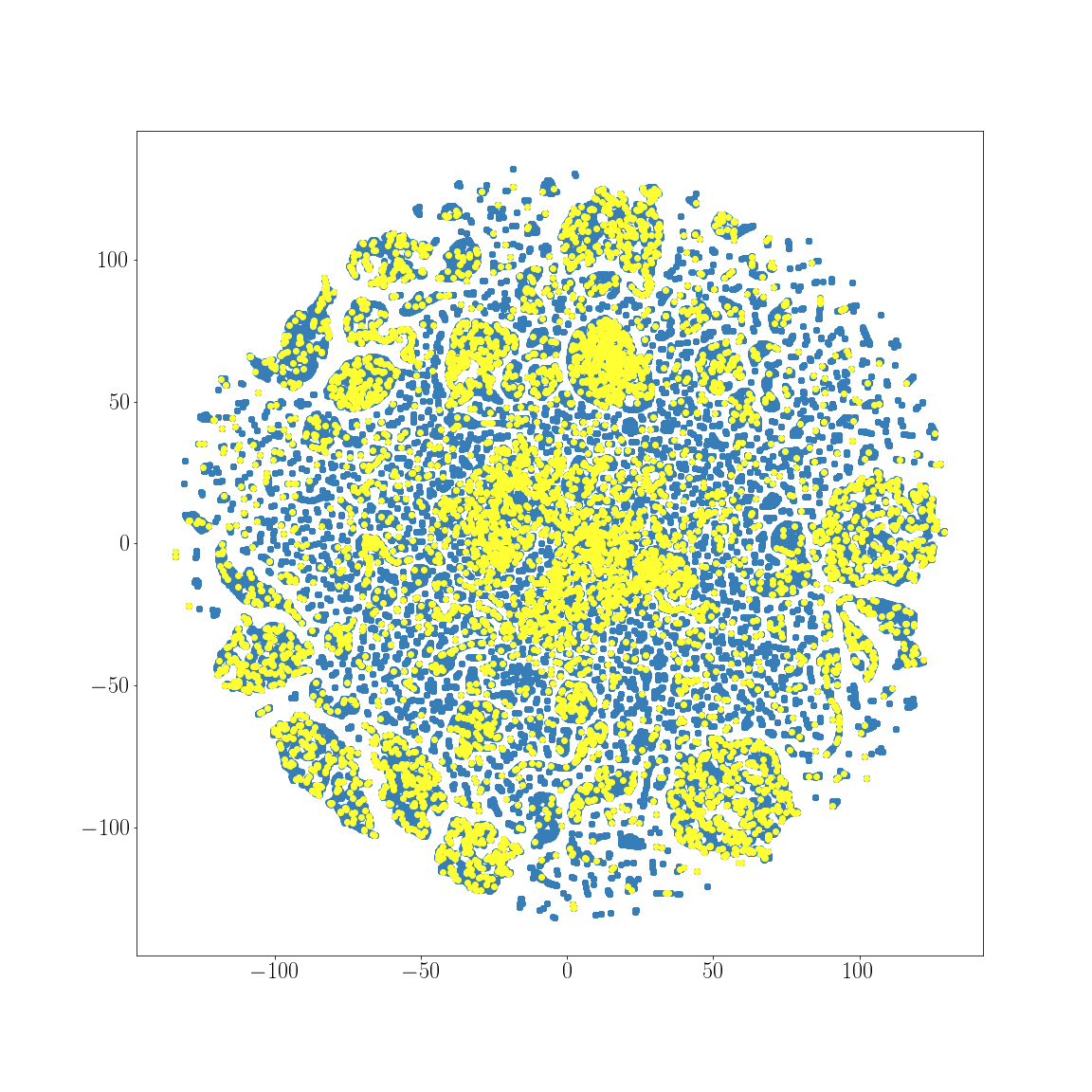}
            \label{fig:radius:1}
        }
        \subfigure[Radius$=$0.3]{%
        \includegraphics[width=.45\columnwidth,trim=4cm 4cm 4cm 4cm,clip]{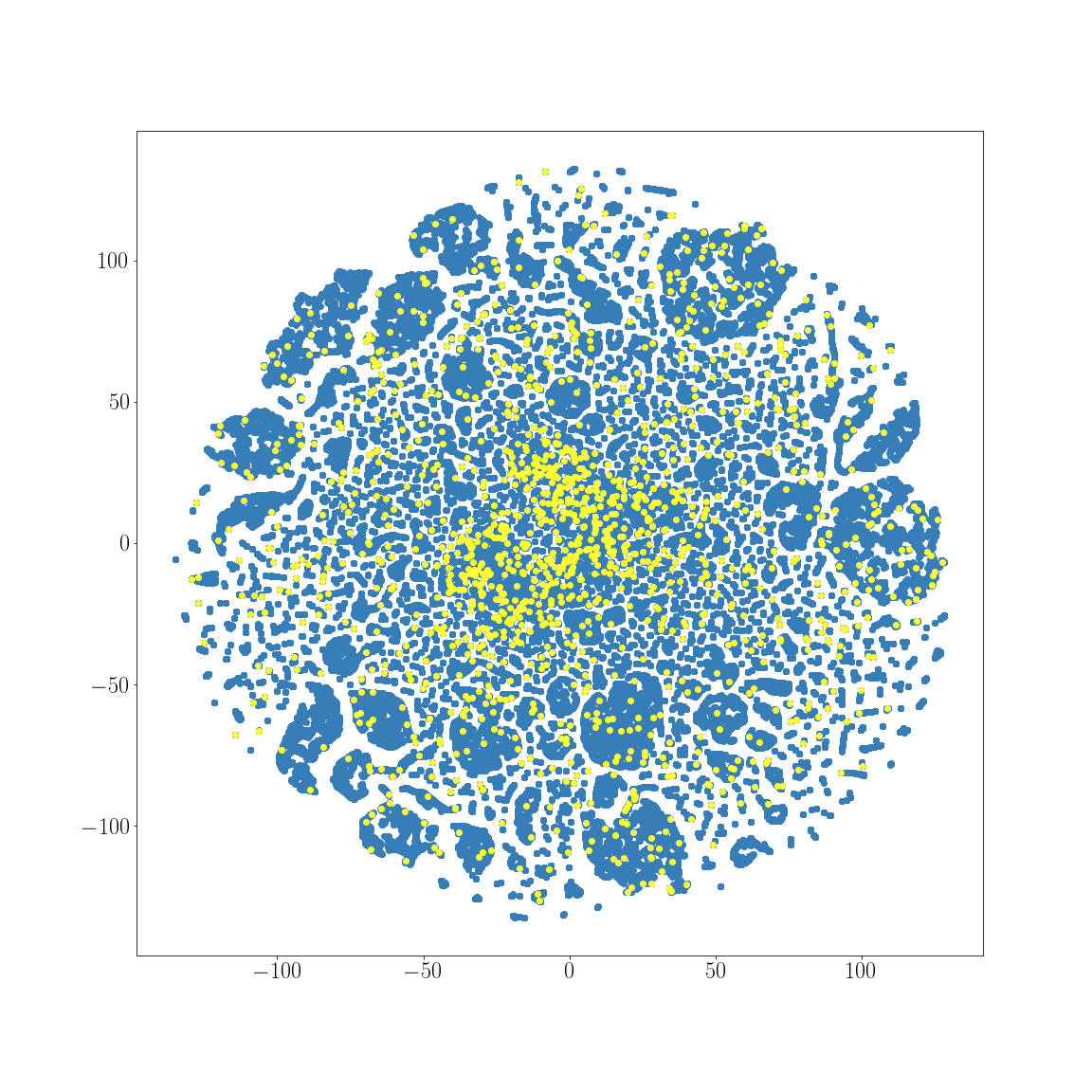}  
        \label{fig:radius:3}
        }}
    \caption{Coverage-based selection: (a) number of selected RPs as a function of the radius size. (b-c) Illustration of selected (yellow) vs unselected (blue) points for two sample radiuses on the tSNE space (real data). }
\end{figure}

\subsection{Approach}
We devise two downsampling strategies  to respectively (i) preserve user densities or (ii) be inclusive with respect to less populated regions.

\paragraph*{Density-preserving}
It is immediate to realize that simple \emph{uniform random selection} preserves user densities: as we expect users to more frequently visit some portion of the physical space, 
random sampling across all STA samples will \emph{preserve user densities}.
The advantage of this approach is that (i) it is trivial to implement, (ii) it favors 
areas that users are frequently visiting: as such, WLAN operators can expect an immediate benefit as measured by 
\elevenk\ reports.  The downside is that it is less resilient to (i) changes in user habits/environment, as well as to (ii)  wrong choices of the sampling period.

\paragraph*{Coverage-friendly}
While it is clear that more radio resources should be made available to denser areas, coverage needs to be secured across the network also for less populated regions in order to provide sufficient service there when needed -- which density-preserving selection cannot guarantee.
To counter this, we additionally propose a \emph{stratified random selection} strategy, whose goal is to enlarge the selection of RPs to not only cover frequently visited areas, but to also ensure
selection of areas that are more rarely visited, in order to avoid coverage holes -- which is thus  \emph{coverage-friendly}.
This strategy is more involved and comprises two steps. First,  we 
project \elevenk\ samples from the PL space with $|\AP|$ dimensions into a lower three-dimensional space, where densities are easier to measure. Specifically, we make use  of t-SNE~\cite{van2008visualizing}, initialized with Principal Components Analysis (PCA), using a perplexity of 40 and 2500 iterations, though we point out that alternative projection methods may fit the purpose. Second, we select an RP in the space at random, and discard all points within a given radius. Specifically, we use k-d tree\cite{bentley1975multidimensional} to ease the search of neighbors within a radius $r$ around its t-SNE projection.
The selection continues until all points are either selected or discarded. 
Notice that whereas the selection process is computationally costlier than the random selection, it needs to be performed infrequently.

\begin{figure}[t]
        {\centering
        \subfigure[UA coverage-friendly]{%
            \includegraphics[width=.45\columnwidth]{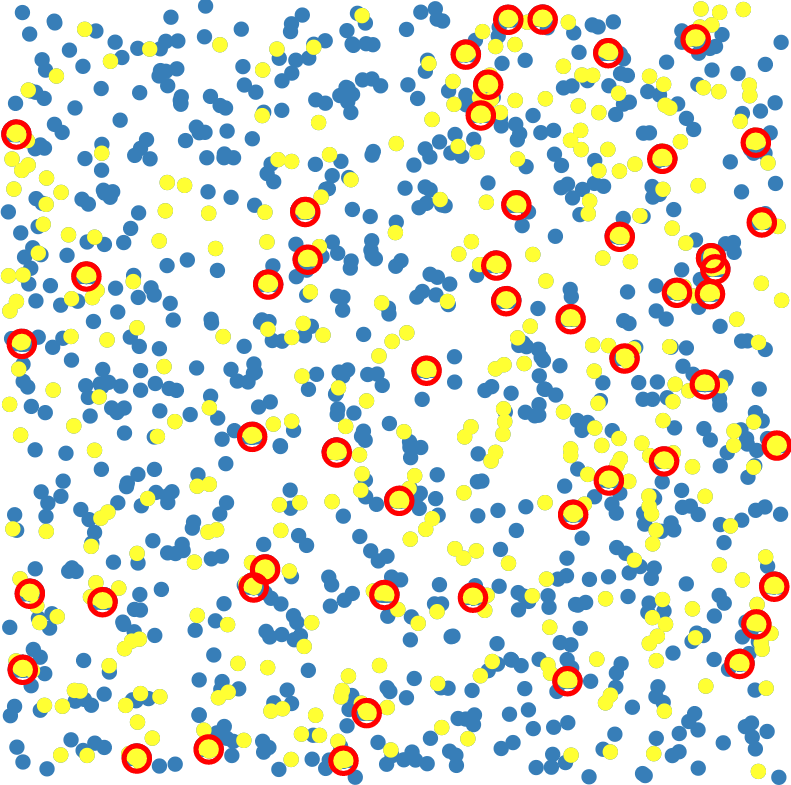}
            \label{fig:cluster_selection}
        }
        \subfigure[UA density-preserving]{%
        \includegraphics[width=.45\columnwidth]{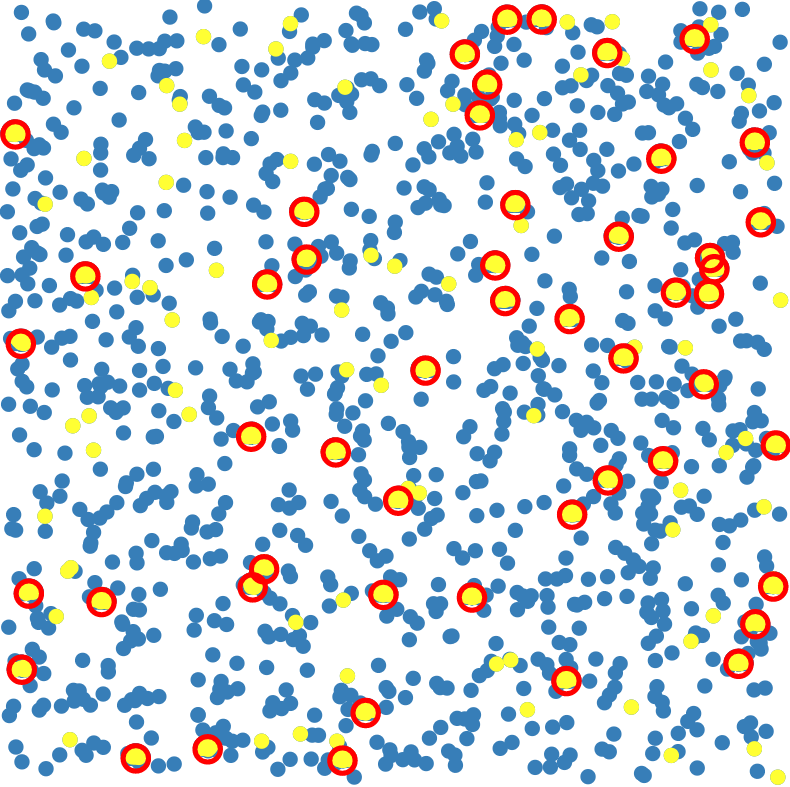}  
        \label{fig:random_selection}
        }}
    \subfigure[Real network densities]{
    {
    \begin{annotatedFigure}
{\includegraphics[width=1.\linewidth]{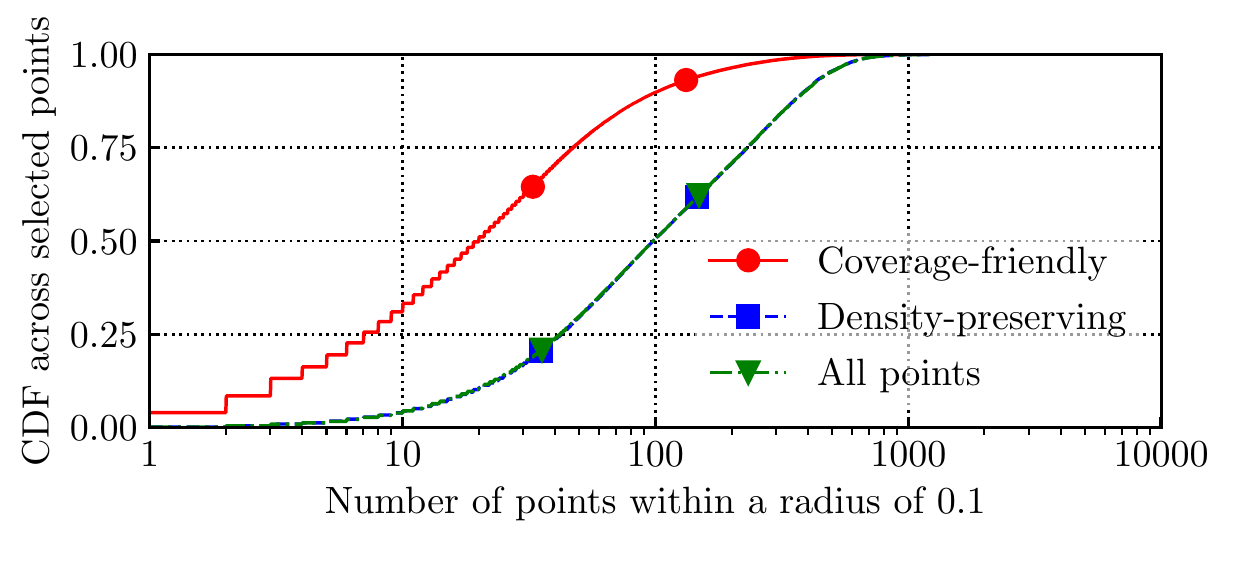}}\label{fig:neigh:selected}
\annotatedFigureText{0.133,0.68}{red}{0.30}{Isolated RPs\\ \textit{Coverage-friendly yields $0.6\%$ better util}}
\annotatedFigureText{0.64,0.59}{blue}{0.28}{Dense RPs\\ \textit{Density-preserving yields $7.3\%$ better util} }
\end{annotatedFigure}
    }\label{fig:wfh}}
    \caption{Comparison of coverage-friendly and density-preserving selections: (a-b)  synthetic scenario with 50 hotspots in a 2-d grid, (c) impact on real network.}
    
\end{figure}

\subsection{Evaluation}

\paragraph*{Complexity viewpoint}
Unlike the more trivial random uniform selection, stratified sampling depends on the chosen radius $r$ as we show in Fig.~\ref{fig:radius:impact}. 
From a computational viewpoint, the interesting regime is for $r\in[0.08, 0.4]$ where, for our 2-radio WLAN, the number of RPs is between 100k and 10k respectively (out of a total of 1.8M), and consequently  the optimization process is expected to last 1.5hr-10min.
Fig.~\ref{fig:radius:1}-\ref{fig:radius:3} illustrate the same effect spatially on the 2-radio dataset, for two example radiuses.  Overall more (and specifically more isolated) points are selected for smaller radius. From here on out, to balance complexity vs selection granularity, we fix the radius to $r=0.1$ (corresponding to a 1hr LS runtime). 

\paragraph*{Representativeness viewpoint}
Next, we analyze the effect of the two strategies described above when they are applied in our power optimization pipeline.
First, we qualitatively observe in  Fig.~\ref{fig:cluster_selection}-(b) the 
RP selection strategies in a simulated physical space with 10k STA positions (blue points), 90\% of which are tightly clustered  around 50 hotspots (red circles). Around 10\% RPs (yellow points) are then chosen, in Fig.~\ref{fig:cluster_selection}
 with the coverage-friendly   and 
 in Fig.~\ref{fig:random_selection}  with the 
density-preserving 
 strategies: as intended, the former selects considerably more RPs outside the hotspots.

For the real 2-radio \elevenk\ scenario, Fig.~\ref{fig:wfh} depicts the distributions of the number of neighbors per point in the projection space.
First, as expected, the distributions of density-preserving points and all points are overlapping.
Second, around 25\% of coverage-friendly selection RPs are isolated (7 or less neighbors, red shading), compared to less than 4\% for density-preserving. 
Third, 25\% of points are in dense areas (230 or more neighbors, blue shading) under density-preserving, as opposed to 3\% only for coverage-friendly. Thus as expected, coverage-friendly oversamples rarer points. Finally, we annotate shaded zones with relative utility difference, gathering confirmation that coverage-friendly slightly favors isolated points (by $0.6\%$), without drastically reducing utility of points in dense areas (favored by $7.3\%$ under density-preserving strategy), which preliminarily supports the interest of these complementary strategies.

\newcommand{\noapprox}[0]{$\approx$} %
\renewcommand{\noapprox}[0]{} 

\begin{figure*}[t]
         \includegraphics[width=0.48\textwidth]{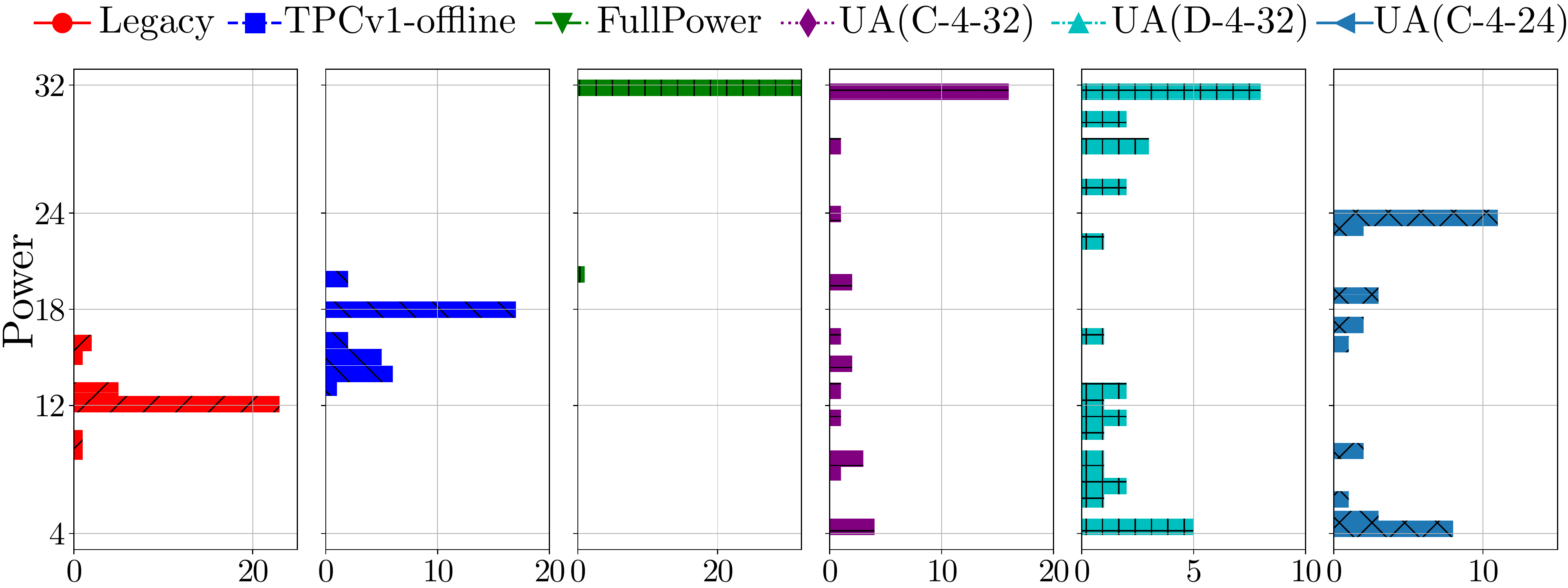} 
        \includegraphics[width=0.51\textwidth,trim=0cm 1.2cm 0cm 0cm,clip]{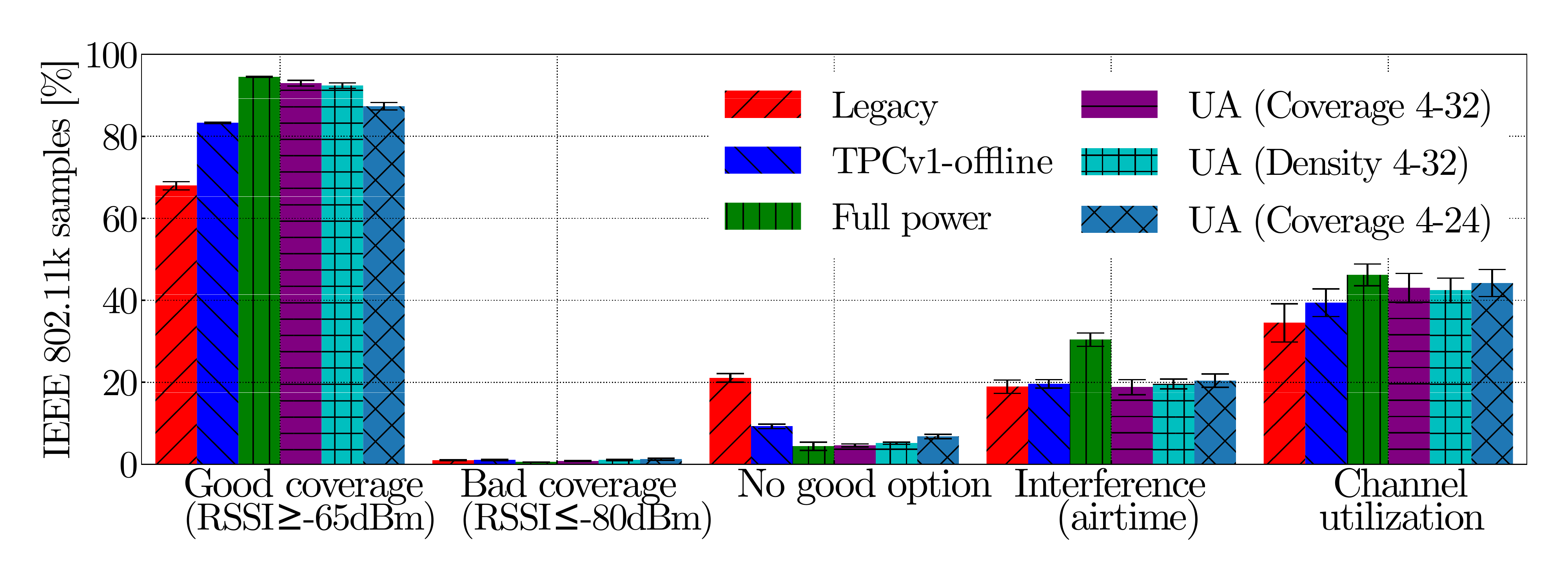}\\  
        \hspace*{130pt}\small{(a)}\hspace*{230pt}\small{(b)}\\
    \small
  \begin{tabular}{lccccccc}
\toprule
\textbf{Strategy}$^\dagger$   & \textbf{\elevenk\ samples} & $\mathbf{\EX(p)}$  &  \textbf{Load}$^\star$\textbf{[\%]}  & \textbf{Interf}$^\star$\textbf{[\%]} & \textbf{DL RSSI}$^\star$\textbf{[dBm]} & \textbf{UL RSSI}$^\star$\textbf{[dBm]} & \textbf{Cvg}$^\ddagger$ \textbf{G/B/O [\%]}  \\
\midrule
Legacy  & 4.9M   & {\bf 12.4}  &  3/10/22 & \textbf{10/18}/31 & -67/-63/-59 & \textbf{-58/-53/-49} & 68/0.9/21   \\ 
\TPCvOne  &   \noapprox 5.8M & 16.7 &   4/11/23 & \underline{11/20}/33 & -64/-59/-55 & \underline{-60/-54/-50}  & 83/1.0/9\\    
Full power  &   \noapprox 4.9M  & 31.6 &    4/10/18 & 19/30/42 & \textbf{-54/-47/-42} & -63/-57/-51 & {\bf 95/0.5/4} \\ 
UA (Coverage-4-32)   &   \noapprox 6.1M & 24.8 &    \textbf{7/20/30} & \textbf{10/18/28} & \underline{-58/-51}/-46 & -64/-58/-52 & \underline{93/0.8/4}\\    
 UA (Density-4-32) &   \noapprox 5.9M  & 22.0 &   \underline{6/17/28}  & \underline{11/20/29} &  \underline{-58/-51/-45} & -64/-57/-52 & 92/1.0/5 \\   
UA (Coverage-4-24)  &   \noapprox 6.2M  & \underline{16.0} &   \underline{6/17/28} & 11/21/30 & -62/-56/-51 & -62/-56/-51 & 92/1.2/6 \\    
\bottomrule
 \multicolumn{8}{l}{$^\dagger$ \footnotesize{For each metric, results of the best strategy are reported in {\bf bold}  and the second best option  is \underline{underlined}}.}\\
   \multicolumn{8}{l}{$^\star$ \footnotesize{First, second and third quartile statistics reported with Q1/Q2/Q3 notation}}\\
 \multicolumn{8}{l}{$^\ddagger$ \footnotesize{Fraction of samples with Good (RSSI$>$-65dBm) or Bad (RSSI$<$-80dBm) coverage, and fraction of badly covered samples with no alternative option}.} 

\end{tabular}

    \caption{Overview of experimental results: (a) breakdown of power levels across APs induced by different strategies, (b) corresponding aggregated results over the full experimental campaign,     along with tabulated details.}
  \label{fig:result_overview}
\end{figure*}

\section{Results from Real Deployment}\label{sec:results}
\subsection{Experimental settings}

\paragraph*{Environment} To evaluate our approach, we deploy and evaluate a set of candidate solutions in an operational network. As 3-radio networks are not pervasive yet, we test on the 2-radio network mentioned in Tab.~\ref{tab:datasets}, comprising 33 APs used by several thousand unique STAs per day. Additionally, since imputation in 2-radio networks is both more needed and more challenging than in 3-radio (cfr. Sec.\ref{sec:imputation}),  we expect this deployment to yield  conservative performance results. 

\paragraph*{Implemented solutions}
We consider a stack of competing solutions: for each we apply the network configuration once, and collect performance measurement  from \elevenk\ STA statistics.
Experiments were carried out between Sep'21 and Jan'22 during which network users   consumed one order of magnitude more data on downlink than uplink. To smooth out statistical variations due to differences in load, user behavior, etc., 
each experiment runs for an entire week. 
Our candidate solutions are: (i) the default product  configuration in the managed network, denoted as \emph{Legacy}; 
(ii)  the  state of the art solution, represented by \emph{TPCv1}~\cite{cisco_v1};
(iii) for reference, a na\"ive \emph{FullPower}\footnote{Notice that all but one APs are capable of setting power level to  32 dBm.} solution expected to maximize both  received signal and  interference;
(iv)-(vi) three configurations of our proposed user-aware system, denoted as \emph{UA}.

As to TPCv1, the original Cisco implementation happens online, i.e., changes are actuated directly in the network in an incremental fashion. To perform a direct comparison among strategies,  
we re-implement the same algorithm, but evaluate its steps offline via our simulator, and only apply the final resulting configuration, denoted as \emph{\TPCvOne}. 
This ensures the comparison  not to be biased due to transient effects (e.g., the slow convergence process of the iterative procedure). 
 
For UA, we implement 3 configurations, where  50k RPs (out of 1.8M samples with $\geq 3$ measured PLs) are selected once in a density-preserving and once in a coverage-friendly way.  We then either use the full scale of power levels 4--32\,dBm, or restrain to 4--24\,dBm, aiming to achieve different operational points in the coverage vs interference tradeoffs.


\subsection{Results at a glance}
Results from the real deployment are succinctly reported in Fig.~\ref{fig:result_overview}.
The top left plot Fig.~\ref{fig:result_overview}-(a) shows the resulting power configuration for each strategy; the top right plot Fig.~\ref{fig:result_overview}-(b) reports the resulting performance under each strategy as measured by a week of \noieee \elevenk\  samples;
results are additionally tabulated and, for  readability, each metric's best strategy is highlighted in bold (the second best is underlined).

\begin{figure*}[t!]
    \subfigure[]{%
        \includegraphics[width=1.25\columnwidth]{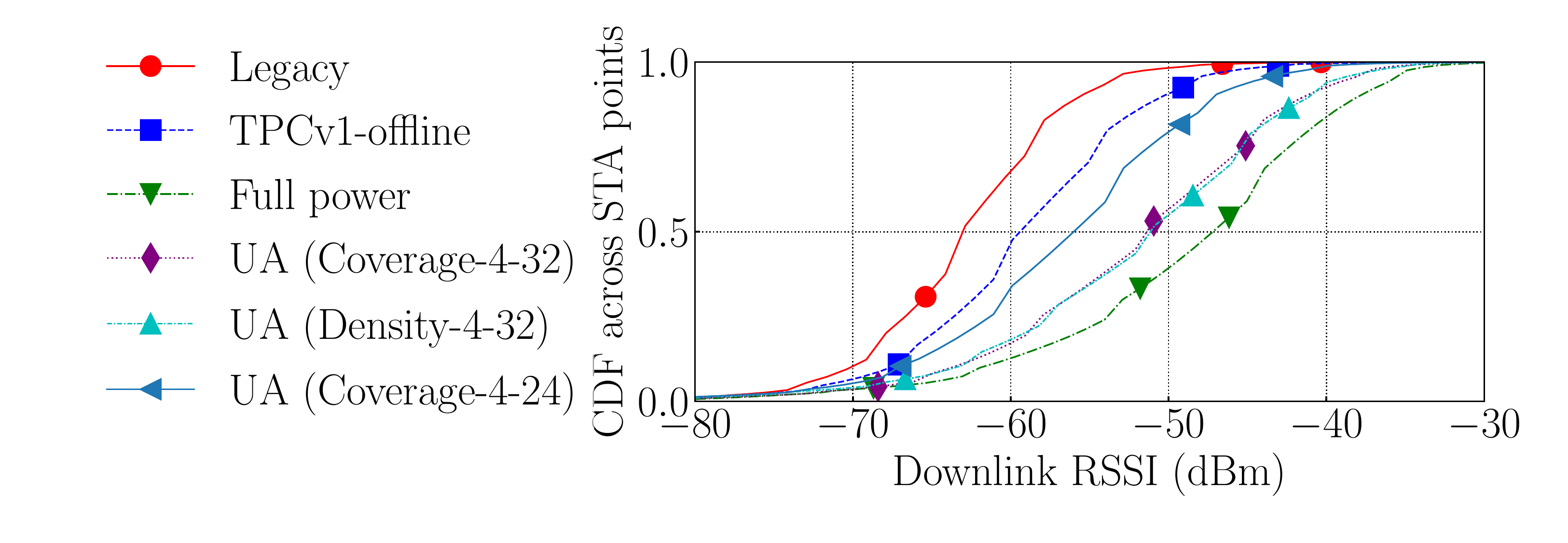}  
        \label{fig:dl}
    }
    \subfigure[]{%
        \includegraphics[width=0.75\columnwidth]{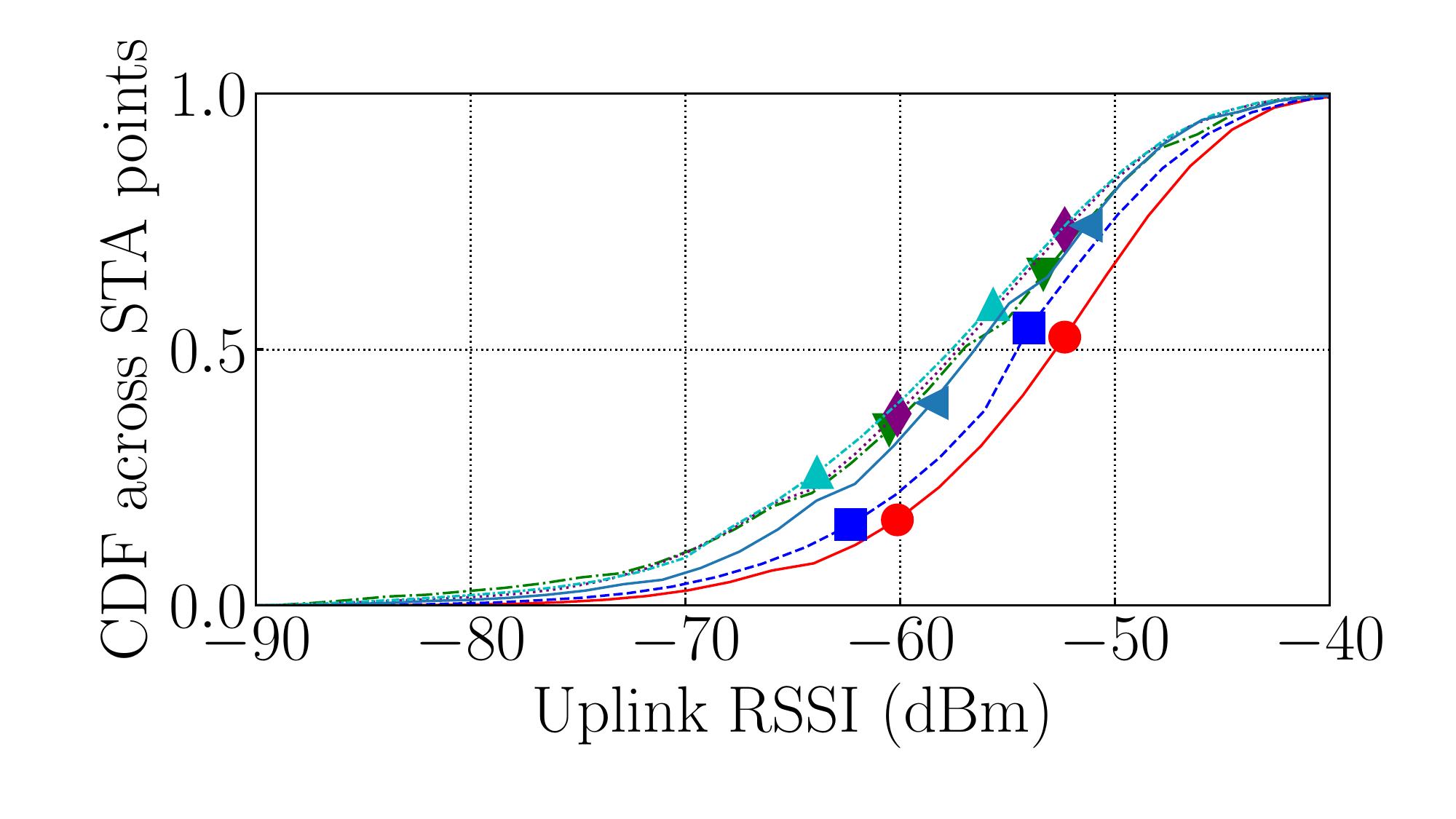}  
        \label{fig:ul}
    }
    \subfigure[]{%
        \includegraphics[width=.65\columnwidth]{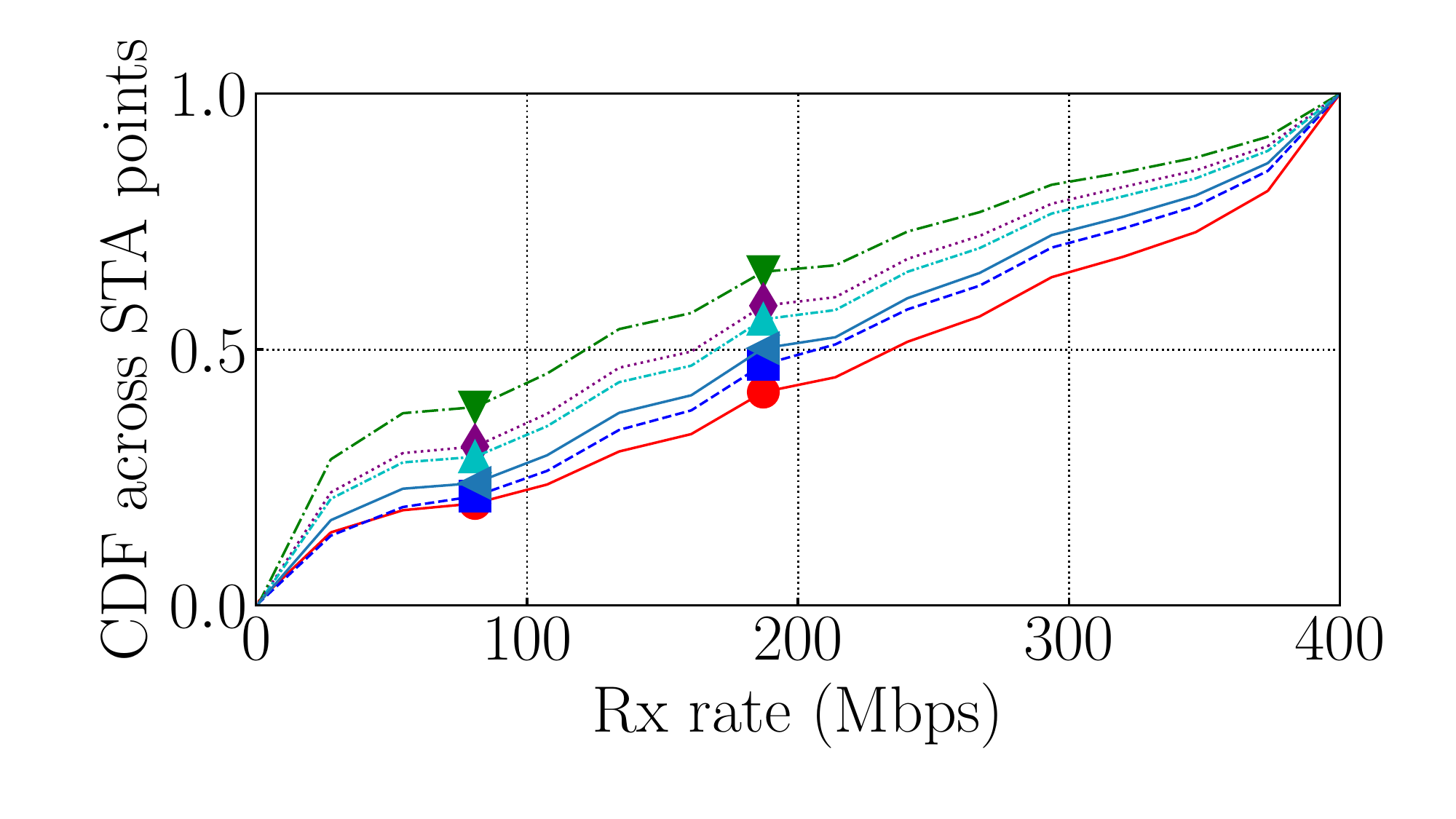}  
        \label{fig:sta:rxrate}
    }
    \subfigure[]{%
        \includegraphics[width=0.65\columnwidth]{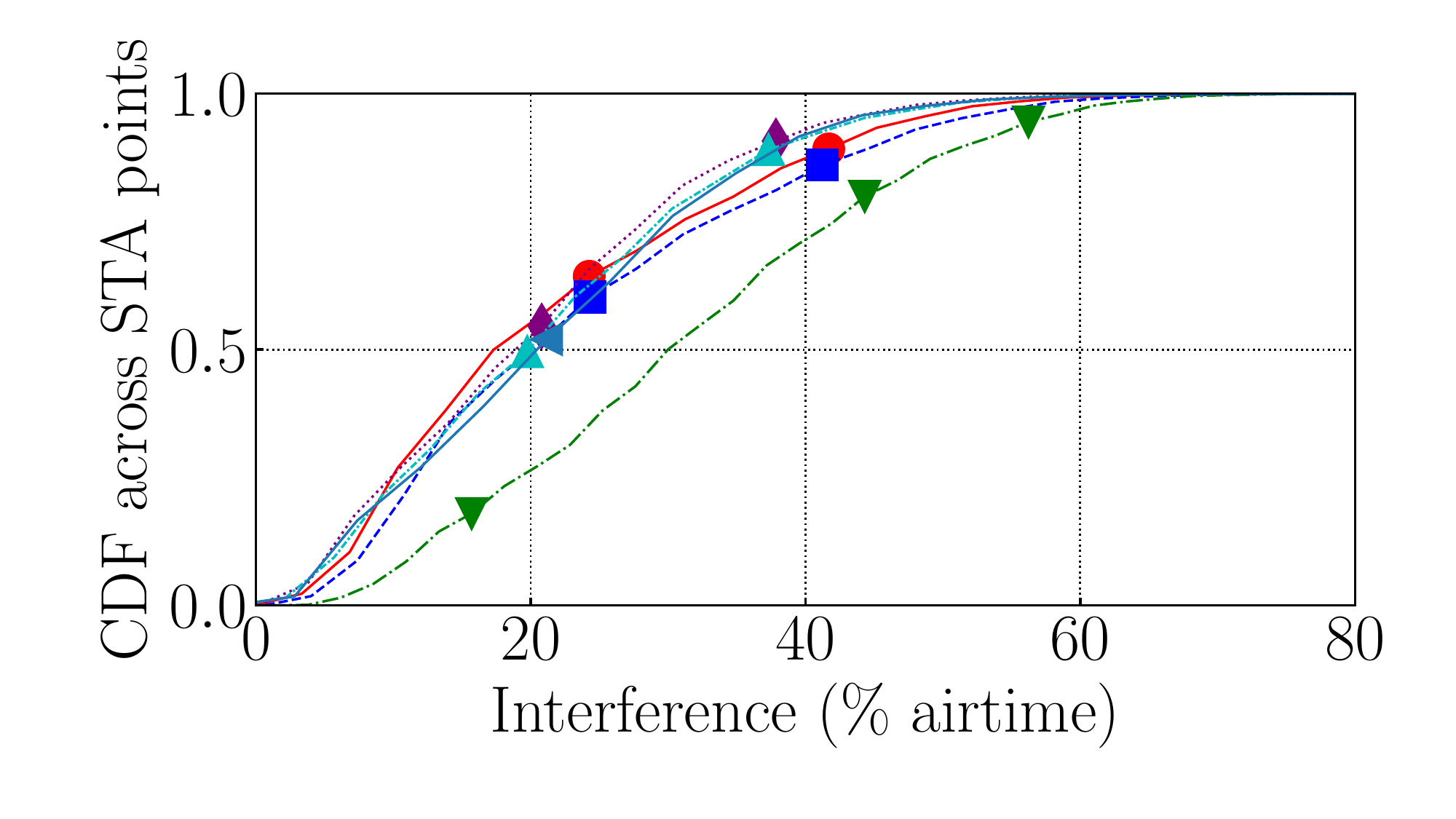}  
        \label{fig:interference:cdf}
    }
    \subfigure[]{
        \includegraphics[width=.65\columnwidth]{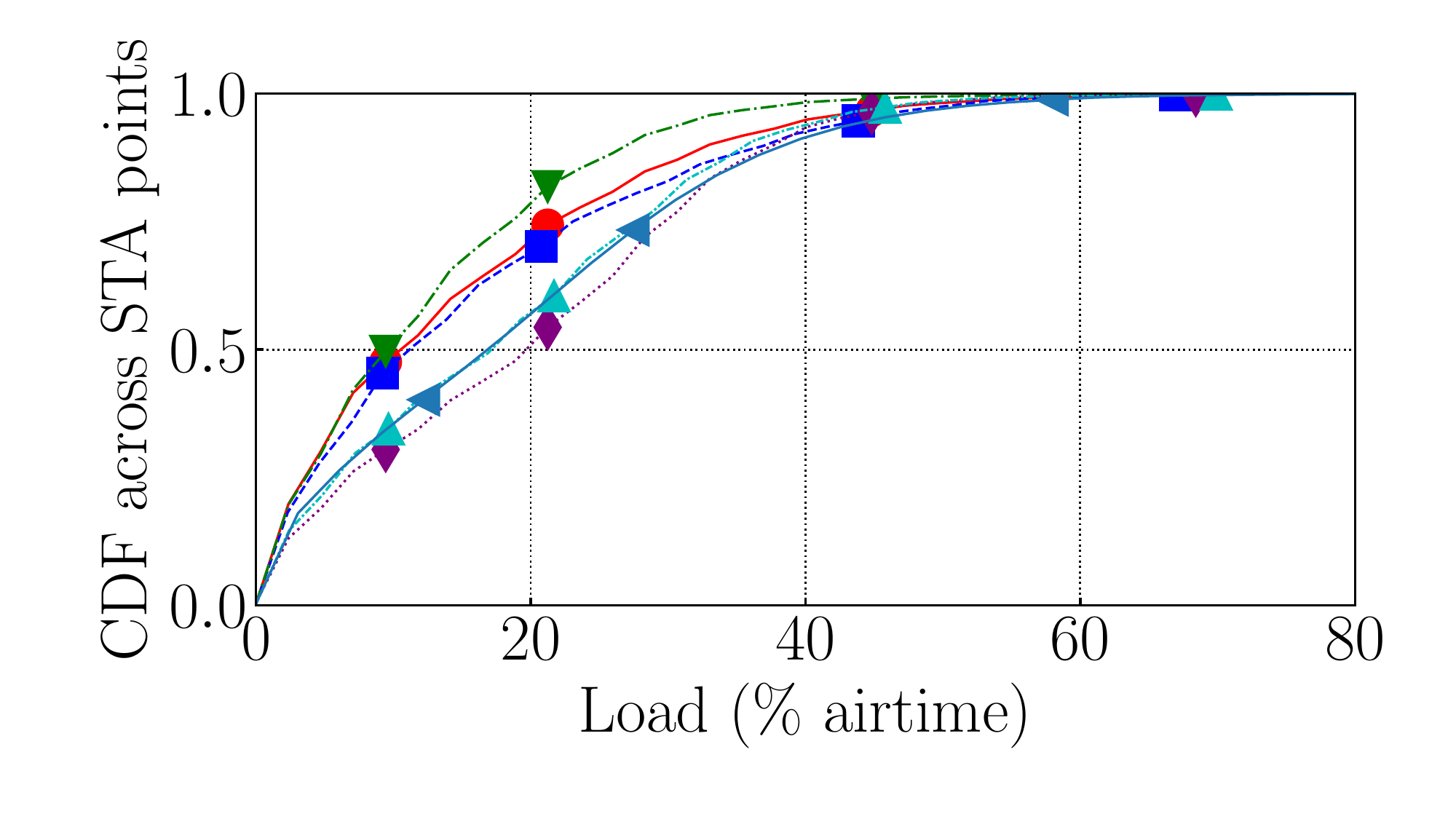}
        \label{fig:load}
    }
    \caption{Detailed experimental results: distribution of \noieee \elevenk\ STA measurements of (a) downlink RSSI,  (b) Uplink RSSI, (c) receive rate (d) percentage of airtime interference from other STA (e) percentage of useful STA airtime.}
  \label{fig:cdf}
\end{figure*}

\paragraph*{Output power configurations}
Examining power configuration first in Fig.~\ref{fig:result_overview}-(a),  we observe an interesting macroscopic difference in the structure of the solutions: i.e.,  Legacy, \TPCvOne\ and FullPower solutions exhibit a unimodal power distribution, in the sense that power levels are tightly centered around a dominant value (12 for Legacy, 18 for TPCv1 and 32 for FullPower). Conversely, all three user-aware strategies have a multi-modal distribution, with two prominent modes toward the extreme of the power level ranges. This is interesting as the UA solutions appear to exploit power diversity, adjusting to measured STA densities. 


\paragraph*{Overall performance}
  Fig.~\ref{fig:result_overview}-(b) reports the most important metrics (daily average and confidence intervals).
  We see that UA solutions increase the fraction of \noieee \elevenk\ samples with good coverage (i.e., where a STA reports at least -65\,dBm of RSSI)
  to roughly the same level of FullPower, but in contrast to the latter, without significant increase of interference (measured as percentage of airtime).  Additionally, with UA solutions,  STAs rarely report coverage issues:  the fraction of readings with bad coverage (RSSI below -80dBm) remains low, and the fraction of STAs that do not have a viable good coverage option (i.e., where no neighbor AP has an RSSI above -65dBm) also  reduces significantly with respect to Legacy. Finally, channel utilization can be seen as an indicator of the network load -- observe that the result comparison is fair. 

\paragraph*{Summary}
From the tabulated results (that detail quartiles for some metrics), it is interesting to notice that while Legacy is the most parsimonious solution (12.4 dBm average), the UA Coverage 4-32 achieves a lower interference despite a higher average power (24.8 dBm). Compared to FullPower, UA solutions reach close to best RSSI and coverage results at  lower interference. The table also  shows  a downside of UA solutions: the increased power at some APs increases the asymmetry between uplink/downlink RSSI (as to FullPower). This is avoided in Legacy and \TPCvOne\ due to lower and more homogeneous power levels.  Observe that UA Coverage 4-24 mitigates this effect.
Interestingly, the coverage-friendly strategy is empirically shown not to negatively affect performance: this could be due to the large volume of samples in dense areas, 
even when aggressively downsampled. 

\subsection{Detailed performance statistics}

We finally provide more in-depth analysis by detailing performance statistics 
shown as empirical cumulative distributions in Fig.~\ref{fig:cdf}.   Generally, this more detailed view of measurement statistics confirms the analysis of the previous section. 
 
\paragraph*{Downlink/uplink RSSI and receive rate}
Distributions for  downlink/uplink RSSI are reported in Fig.~\ref{fig:cdf}-(a,b): CDFs are clearly separated in the case of downlink RSSI, with an order of increasing performance for Legacy, TCPv1, UA strategies toward FullPower. The coverage-friendly UA strategy is however indistinguishable, from a downlink RSSI perspective, from the density-preserving one (when both on 4--32 dBm).
This enhancement in downlink comes with a (smaller) degradation in uplink RSSI. Fig.~\ref{fig:sta:rxrate} confirms this finding, observing an impact on the AP rx rate\footnote{Rate during transmission excluding interference waiting time}. This is probably due to STAs that do not increase their power levels symmetrically to APs. UA strategy with less power (4--24) is a compromise that achieves an rx rate comparable to TPCv1.
Overall, however, the negative impact on the uplink is weaker (e.g. -5dBm RSSI in the median w.r.t.\ Legacy) than the benefits to downlink RSSI statistics (+15 dBm in the median  w.r.t.\ Legacy).

\paragraph*{Interference and load}
Looking at airtime interference reported in Fig.~\ref{fig:interference:cdf}: For UA solutions, only the worst 10\% samples have an interference in excess of 40\%; it increases to 15\% for \TPCvOne, and to 30\%  for FullPower.
Note that interference here also includes 
waiting time during transmissions by the same AP to other associated STAs. 

Finally, we see in Fig.~\ref{fig:load} the distributions of load across samples, where load is defined as the percentage of airtime used by the STA for transmission. Higher values here represents less waiting time through interference, and, assuming all other  conditions being equal, higher throughput. As the families of curves are clearly separated, we  see that the group of UA results improves over Legacy and \TPCvOne, while FullPower achieves the least favorable result. The fact that the measured load is high for the UA family suggests that  STA load is well balanced across APs under UA strategies.

\paragraph*{Summary} Results from our 
deployment 
show that our UA approach improves WLAN performance, increasing the downlink signal, decreasing interference and  balancing load (indirectly shown through increased STA load). Potential issues with lower uplink signal can be (if deemed serious) mitigated  by an appropriate range of  allowed power levels. 
Finally, we see that coverage-friendly allocation, designed for 
coverage across all zones including the least popular ones, does not negatively affect denser areas in any  performance metric.

\section{Prior work}\label{sec:related}
Related work to ours concerns either (i)  WLAN transmit power optimization or (ii) user-aware WLAN management. 

\paragraph*{Power optimization} Prior work studied WLAN power optimization aiming at different objectives and applying various techniques. Some of the papers that we mention assume, as a simplification, knowledge about user positions or channel quality to the serving AP but, unless mentioned explicitly, they do not consider the practical aspects of leveraging available data for a true user-aware WLAN power optimization.
A classic approach is called  \emph{cell breathing}  \cite{bahl2006cell, 1543695, 1313270,villegas2006load, garcia2008cooperative} which generally  targets load balancing by ongoing power adjustment.
Closer to our work in terms of \emph{approach},
authors in \cite{kuhn2005interference} tune  power  to minimize the largest number of covering APs at any STA, using classic optimization techniques with simplistic STA and interference modeling. Joint channel and simple power configuration (APs on/off) are
considered in \cite{8240655}, where   
heuristics are applied for  interference minimization, and STAs are  modeled as connected regions. These papers share the  same spirit, but significantly differ in the extent of validation -- while we make simplifying assumptions in the design phase, our validation phase is carried out in a real large scale operational network. Two works that assume perfect channel state knowledge are \cite{tang2014joint} which approximates optimal power and transmit rate, and \cite{eisen2019large}, where the authors employ a Graph Neural Network for power optimization. Minimizing power consumption while guaranteeing Quality of Service, \cite{lee2016just} assumes full knowledge as well, and even control over STA association. Closer to our work in terms of \emph{objective} are \cite{zhao2019:DRL2,10.1145/2695664.2695718}. In particular,  the objective in~\cite{zhao2019:DRL2} is to maximize channel capacity, employing Reinforcement Learning (RL) for state exploration, though  STAs are simulated with synthetic densities. Conversely, \cite{10.1145/2695664.2695718} aims at minimizing interference under coverage constraints, reflecting users as virtual test points, and using a simple pathloss formula to estimate coverage. 
Overall, most of the work on AP power optimization was tested in simulation and small testbeds (e.g.\ 8 APs in \cite{8240655}, while \cite{10.1145/2695664.2695718}  employs 7 APs out of 24 available in an operational network).


\paragraph*{User-aware WLAN management} A classic user-aware task in the context of WLAN is user localization, which may indirectly provide valuable insights into WLAN resource management.
In particular \elevenk\ RSSI measurements are used for distance  estimation~\cite{zanella2016best} or localization~\cite{liu2021wibeacon,koo2010localizing,zhao2014rssi, khalajmehrabadi2016joint, chen2017awl, rea2019smartphone}, in the case of \cite{ayyalasomayajula2020deep} leveraging convolutional neural networks. Our approach, applied to power management, is to avoid the distance (or position) estimation altogether, and work from density of an embedding space directly as a proxy of the density in the real space. 
More directly applied to WLAN resource management:
some proposals ~\cite{villegas2006load, garcia2008cooperative} exploit \noieee \elevenk\ data (to achieve a favorable STA association), 
  and are thus related to ours. In contrast to the latter work (where the resulting power configuration is dependent on the current status) our approach aims at a global optimum. Most importantly, the proposed algorithms assume complete \noieee \elevenk\ reports -- whereas we find that they are usually  incomplete and their exploitation requires careful design of automated ML tools. Recent work~\cite{gavcanin2019self} proposes to leverage user-related RSSI measurements to recommend best positions for WiFi extenders in home environments, showcasing the utility in a testbed.
In contrast, our work is the first to systematically leverage such measurements at scale in an operational network.


\section{Limitations}\label{sec:limit}
While we showed feasibility of user-aware power control and quantified its gains, a number of challenges however remain.

\paragraph*{Explainability}
Given that interference, load and coverage interact with each other in non-trivial ways, and 
that user densities are manipulated in a latent space (and not in the physical one), the decisions of the algorithm are not easy to understand. Here, we formulate a utility function per reference point, we apply it on tens of thousands of reference points, and pick the configuration that leads to the highest estimated utility. This 
approach leads 
to better results in practice, but the rationale behind the chosen decisions needs more investigations.

\paragraph*{Lack of control on terminals}
This is not a limitation of our system per se but one that limits the potential of user-aware AP power control in practice. As we show in Fig.~\ref{fig:ul}, our approach comes at the cost of a slightly worse uplink RSSI. This is not directly due to our approach but to the fact that terminals who adopt a proprietary behavior do not reciprocally increase their power levels when the APs do so. The lack of control on user terminals is a known limitation of WLAN networks in general, unlike cellular ones. 

\paragraph*{Historical user-density update}
Our approach optimizes for user densities of presence inferred from historical data. In our experiments, we used reference points from May to June and assumed that the resulting densities remained valid during our test campaign from September to January. 
This worked for us because the network, its corresponding building configuration and users did not drastically change inbetween. This might not be the case in other settings. More work is needed to understand which events impact the stability of user patterns and, if necessary, the rate at which RPs should be updated.

\section{Conclusions}\label{sec:conclusions}
We presented and evaluated in a production environment, the first user-aware WLAN AP transmit power control system: results show a +15dBm improvement in median signal strength, with no impact on interference, which on the contrary decreases. 
Along the way, we contributed (i) a novel method to estimate the performance experienced by mobile terminals based on their position in the pathloss space (as opposed to the less accurate physical one), and a (ii) novel ML-based method to impute missing signal strength measurements.
This makes our system complementary to any channel or bandwidth allocation strategy which could be further used to optimize the network following our user-aware utility. We outlined some limitations which could be tackled in future work.
\bibliographystyle{plain}
\bibliography{reference}

\begin{thebibliography}{10}

\bibitem{ahmed2006successive}
Nabeel Ahmed and Srinivasan Keshav.
\newblock A successive refinement approach to wireless infrastructure network
  deployment.
\newblock In {\em IEEE WCNC 2006}, pages 511--519, 2006.

\bibitem{akella2005self}
Aditya Akella, Glenn Judd, Srinivasan Seshan, and Peter Steenkiste.
\newblock Self-management in chaotic wireless deployments.
\newblock In {\em ACM MobiCom 2005}, pages 185--199, 2005.

\bibitem{ayyalasomayajula2020deep}
Roshan Ayyalasomayajula, Aditya Arun, Chenfeng Wu, Sanatan Sharma,
  Abhishek~Rajkumar Sethi, Deepak Vasisht, and Dinesh Bharadia.
\newblock Deep learning based wireless localization for indoor navigation.
\newblock In {\em ACM MobiCom 2020}, pages 1--14, 2020.

\bibitem{bahl2006cell}
Paramvir Bahl, Mohammad~T Hajiaghayi, Kamal Jain, Sayyed~Vahab Mirrokni, Lili
  Qiu, and Amin Saberi.
\newblock Cell breathing in wireless {LANs}: Algorithms and evaluation.
\newblock {\em IEEE Trans. Mob. Comput.}, 6(2):164--178, 2006.

\bibitem{bentley1975multidimensional}
Jon~Louis Bentley.
\newblock Multidimensional binary search trees used for associative searching.
\newblock {\em Communications of the ACM}, 18(9):509--517, 1975.

\bibitem{10.1145/3131365.3131398}
Apurv Bhartia, Bo~Chen, Feng Wang, Derrick Pallas, Raluca Musaloiu-E, Ted
  Tsung-Te Lai, and Hao Ma.
\newblock Measurement-based, practical techniques to improve 802.11ac
  performance.
\newblock In {\em AMC IMC 2017}, 2017.

\bibitem{1543695}
Olivia Brickley, Susan Rea, and Dirk Pesch.
\newblock Load balancing for {QoS} optimisation in wireless {LANs} utilising
  advanced cell breathing techniques.
\newblock In {\em IEEE Vehicular Technology Conference 2005}, 2005.

\bibitem{broustis2009measurement}
Ioannis Broustis, Konstantina Papagiannaki, Srikanth~V Krishnamurthy, Michalis
  Faloutsos, and Vivek~P Mhatre.
\newblock Measurement-driven guidelines for 802.11 {WLAN} design.
\newblock {\em IEEE/ACM Trans. Netw.}, 18(3):722--735, 2009.

\bibitem{cai2010singular}
Jian-Feng Cai, Emmanuel~J Cand{\`e}s, and Zuowei Shen.
\newblock A singular value thresholding algorithm for matrix completion.
\newblock {\em SIAM Journal on optimization}, 20(4):1956--1982, 2010.

\bibitem{chen2017awl}
Zhe Chen, Zhongmin Li, Xu~Zhang, Guorong Zhu, Yuedong Xu, Jie Xiong, and Xin
  Wang.
\newblock {AWL}: Turning spatial aliasing from foe to friend for accurate
  {WiFi} localization.
\newblock In {\em CoNEXT 2017}, pages 238--250, 2017.

\bibitem{cisco_v1}
Cisco.
\newblock Cisco {Wireless} {LAN} {Controller} {Configuration} {Guide},
  {Release} 7.2, 2022.

\bibitem{dong2018pcc}
Mo~Dong, Tong Meng, Doron Zarchy, Engin Arslan, Yossi Gilad, Brighten Godfrey,
  and Michael Schapira.
\newblock {PCC} vivace: Online-learning congestion control.
\newblock In {\em USENIX NSDI 2018}, pages 343--356, 2018.

\bibitem{eisen2019large}
Mark Eisen and Alejandro Ribeiro.
\newblock Large scale wireless power allocation with graph neural networks.
\newblock In {\em IEEE SPAWC 2019}, pages 1--5, 2019.

\bibitem{gavcanin2019self}
Haris Ga{\v{c}}anin, Erma Perenda, Samurdhi Karunaratne, and Ramy Atawia.
\newblock Self-optimization of wireless systems with knowledge management: An
  artificial intelligence approach.
\newblock {\em IEEE Trans. Veh. Technol.}, 68(10):9682--9697, 2019.

\bibitem{garcia2008cooperative}
Eduard Garcia, Rafael Vidal, and Josep Paradells.
\newblock Cooperative load balancing in {IEEE} 802.11 networks with cell
  breathing.
\newblock In {\em IEEE ISCC 2008}, pages 1133--1140, 2008.

\bibitem{iacoboaiea2021real}
O.~{Iacoboaiea}, J.~{Krolikowski}, Z.~{Ben Houidi}, and D.~{Rossi}.
\newblock {Real-Time Channel Management in WLANs: Deep Reinforcement Learning
  versus Heuristics}.
\newblock In {\em IFIP Networking}, 2021.

\bibitem{jiang2017unleashing}
Junchen Jiang, Vyas Sekar, Ion Stoica, and Hui Zhang.
\newblock Unleashing the potential of data-driven networking.
\newblock In {\em Springer COMSNETS 2017}, pages 110--126, 2017.

\bibitem{jiang2017pytheas}
Junchen Jiang, Shijie Sun, Vyas Sekar, and Hui Zhang.
\newblock Pytheas: Enabling data-driven quality of experience optimization
  using group-based exploration-exploitation.
\newblock In {\em USENIX NSDI 2017}, pages 393--406, 2017.

\bibitem{khalajmehrabadi2016joint}
Ali Khalajmehrabadi, Nikolaos Gatsis, Daniel~J Pack, and David Akopian.
\newblock A joint indoor {WLAN} localization and outlier detection scheme using
  {LASSO} and elastic-net optimization techniques.
\newblock {\em IEEE Trans. Mob. Comput.}, 16(8):2079--2092, 2016.

\bibitem{koo2010localizing}
Jahyoung Koo and Hojung Cha.
\newblock Localizing {WiFi} access points using signal strength.
\newblock {\em IEEE Commun. Lett.}, 15(2):187--189, 2010.

\bibitem{kuhn2005interference}
Fabian Kuhn, Pascal Von~Rickenbach, Roger Wattenhofer, Emo Welzl, and Aaron
  Zollinger.
\newblock Interference in cellular networks: The minimum membership set cover
  problem.
\newblock In {\em International Computing and Combinatorics Conference 2005},
  pages 188--198. Springer, 2005.

\bibitem{lee2016just}
Kimin Lee, Yeonkeun Kim, Seokhyun Kim, Jinwoo Shin, Seungwon Shin, and Song
  Chong.
\newblock Just-in-time {WLANs}: {On-demand} interference-managed {WLAN}
  infrastructures.
\newblock In {\em IEEE INFOCOM 2016}, pages 1--9, 2016.

\bibitem{li2019misgan}
Steven Cheng-Xian Li, Bo~Jiang, and Benjamin Marlin.
\newblock Learning from incomplete data with generative adversarial networks.
\newblock In {\em ICLR 2019}, 2019.

\bibitem{liu2021wibeacon}
Ruofeng Liu, Zhimeng Yin, Wenchao Jiang, and Tian He.
\newblock {WiBeacon}: expanding {BLE} location-based services via wifi.
\newblock In {\em ACM MobiCom 2021}, pages 83--96, 2021.

\bibitem{5506716}
Y.~{Liu}, W.~{Wu}, B.~{Wang}, T.~{He}, S.~{Yi}, and Y.~{Xia}.
\newblock Measurement-based channel management in {WLANs}.
\newblock In {\em IEEE WCNC 2010}, 2010.

\bibitem{naseerconfiganator}
Usama Naseer and Theophilus~A Benson.
\newblock Configanator: A data-driven approach to improving {CDN} performance.
\newblock In {\em USENIX NSDI 2022}, 2022.

\bibitem{rea2019smartphone}
Maurizio Rea, Traian~Emanuel Abrudan, Domenico Giustiniano, Holger Claussen,
  and Veli-Matti Kolmonen.
\newblock Smartphone positioning with radio measurements from a single wifi
  access point.
\newblock In {\em ACM CoNEXT 2019}, pages 200--206, 2019.

\bibitem{10.1145/2695664.2695718}
Marcelo Riedi, Giovanna~G. Basilio, and Marcelo~E. Pellenz.
\newblock Channel and power allocation algorithm to optimize the performance of
  large {WLANs}.
\newblock In {\em ACM SAC 2015}, page 673–679, 2015.

\bibitem{tang2014joint}
Suhua Tang, Hiroyuki Yomo, Akio Hasegawa, Tatsuo Shibata, and Masayoshi Ohashi.
\newblock Joint transmit power control and rate adaptation for wireless {LANs}.
\newblock {\em Springer Wireless personal communications}, 74(2):469--486,
  2014.

\bibitem{van2008visualizing}
Laurens Van~der Maaten and Geoffrey Hinton.
\newblock Visualizing data using {t-SNE}.
\newblock {\em Journal of machine learning research}, 9(11), 2008.

\bibitem{1313270}
H.~Velayos, V.~Aleo, and G.~Karlsson.
\newblock Load balancing in overlapping wireless lan cells.
\newblock In {\em IEEE ICC 2004}, volume~7, pages 3833--3836, 2004.

\bibitem{villegas2006load}
E~Garcia Villegas, R~Vidal Ferre, and J~Paradells Aspas.
\newblock Load balancing in {WLANs} through ieee 802.11 k mechanisms.
\newblock In {\em IEEE ISCC 2006}, pages 844--850, 2006.

\bibitem{zanella2016best}
Andrea Zanella.
\newblock Best practice in {RSS} measurements and ranging.
\newblock {\em IEEE Communications Surveys \& Tutorials}, 18(4):2662--2686,
  2016.

\bibitem{8240655}
Y.~{Zhang}, C.~{Jiang}, J.~{Wang}, Z.~{Han}, J.~{Yuan}, and J.~{Cao}.
\newblock Green {Wi-Fi} management: Implementation on partially overlapped
  channels.
\newblock {\em IEEE Transactions on Green Communications and Networking},
  2(2):346--359, 2018.

\bibitem{zhao2014rssi}
Fang Zhao, Haiyong Luo, Hao Geng, and Qijin Sun.
\newblock An rssi gradient-based ap localization algorithm.
\newblock {\em China Communications}, 11(2):100--108, 2014.

\bibitem{zhao2019:DRL2}
Guofeng Zhao, Yong Li, Chuan Xu, Zhenzhen Han, Yuan Xing, and Shui Yu.
\newblock Joint power control and channel allocation for interference
  mitigation based on reinforcement learning.
\newblock {\em IEEE Access}, 2019.

\end{thebibliography}

\end{document}